\documentclass[notoc,11pt,twoside,preprint,nohyper]{JHEP} 
\usepackage{epsfig}
\usepackage{amssymb}
\newcommand\fverb{\setbox\pippobox=\hbox\bgroup\verb}
\newcommand\fverbdo{\egroup\medskip\noindent%
            \fbox{\unhbox\pippobox}\ }
\newcommand\fverbit{\egroup\item[\fbox{\unhbox\pippobox}]}
\newbox\pippobox

\title{When semantics turns to substance: reformulating QCD analysis of
${\mathbf{F_2^{\gamma}(x,Q^2)}}$}

\author{by J. Ch\'{y}la\thanks{Supported by Grant Agency of ASCR under
 grant No. A1010602}\\
    Institute of Physics, Na Slovance 2, Prague 8, Czech Republic\\
    E-mail: \email{chyla@fzu.cz}}
\received{\today}       
\accepted{\today}       

\preprint{PRA-HEP 99/05}  

\abstract{QCD analysis of $F_2^{\gamma}(x,Q^2)$ is revisited. It
is emphasized that the presence of the inhomogeneous term in the
evolution equations for quark distribution functions of the photon
implies important difference in the way factorization mechanism
works in photon--hadron and photon--photon collisions as compared
to the hadronic ones. Moreover, a careful definitions of the very
concepts of the ``leading order'' and ``next--to--leading order''
QCD analysis of $F_2^{\gamma}$ are needed in order to separate
genuine QCD effects from those of pure QED origin. After
presenting such definitions, I show that all existing allegedly LO,
as well as NLO
analyses of $F_2^{\gamma}(x,Q^2)$ are incomplete. The
source of this incompleteness of the conventional approach is
traced back to the lack of clear identification of QCD effects and
to the misinterpretation of the behaviour of $q^{\gamma}(x,M)$ as
a function of $\alpha_s(M)$. Complete LO and NLO QCD analyses of
$F_2^{\gamma}(x,Q^2)$ are shown to differ
substantially from the conventional ones. Whereas complete NLO
analysis requires the knowledge of two so far uncalculated
quantities, a complete LO one is
currently possible, but compared to the conventional
formulation requires the inclusion of four
known, but in the existing LO analyses unused quantities.
The arguments recently advanced 
in favour of the conventional approach are analyzed and
shown to contain a serious flaw. If corrected, they
actually lend support to my claim.}

\keywords{QCD, perturbation theory, photon structure}

\begin{document}

\maketitle 

\section{Introduction}
Observed from a large distance the photon behaves as a neutral
structureless object governed by the laws of Quantum
Electrodynamics. However, when probed at short distances it
exhibits some properties characteristic of hadrons
\footnote{For recent theoretical and experimental reviews see
\cite{Vogt} and \cite{Stefan,Richard}, respectively.}. This
``photon structure'' is quantified, similarly as in the case of
hadrons, in terms of parton distribution functions (PDF),
satisfying certain evolution equations. Because of the direct
coupling of photons to quark--antiquark pairs these evolution
equations are, contrary to the case of hadrons, inhomogeneous.
This inhomogeneity has important implications for QCD analysis
of $F_2^{\gamma}(x,Q^2)$ and other physical quantities involving
photon in the initial state.

In the previous paper \cite{factor} these implications led me to
the conclusion the all existing NLO QCD analyses of
$F_2^{\gamma}(x,Q^2)$ are incomplete. With the exception of the
authors of the FKP approach \cite{FKP} my claim has been either
ignored of rejected. The inertia of the ``common wisdom''
is enormous. I have therefore appreciated the recent
attempt of A. Vogt \cite{VogtinFreiburg} to present
mathematical arguments in favour of the conventional
approach. I analyze them in Section 4 and show that
they contain a serious flaw. If corrected, Vogt's
arguments actually lend support to my claim.

In the course of discussions of this and related subjects
concerning conventional QCD analysis of photon structure, I have
realized that the main source of the confusion surrounding the
QCD analysis of $F_2^{\gamma}(x,Q^2)$ was the lack of a clear
definition of what ``LO'' and ``NLO'' means in the context of
photonic interactions. The point is that all existing
conventional QCD analyses of $F_2^{\gamma}$ mix purely QED effects,
which are usually quite dominant, with the genuine QCD ones,
which represent mostly small corrections only.
I will therefore start by presenting
my definition of what ``leading'' and
``next--to--leading'' order means for parton distribution
functions of the photon and for $F_2^{\gamma}(x,Q^2)$.
In the next step I will construct explicit solutions of
the inhomogeneous evolution equation including
inhomogeneous as well as homogeneous splitting functions up to
order $\alpha_s^2$. Straightforward analysis of these solutions
shows that my claim in \cite{factor} was actually only partially
correct: not only the existing supposedly NLO analyses are
incomplete, but so are the LO ones! Contrary to the NLO
analysis, which is currently impossible to perform because the
necessary quantities have not yet been calculated, complete LO
QCD analysis of $F_2^{\gamma}(x,Q^2)$ is feasible, but requires
the inclusion of several additional $O(\alpha_s)$ quantities.

The paper is organized as follows. In the next Section basic facts
and notation concerning PDF of the photon are recalled, followed
in Section 3 by the discussion of the properties of the pointlike
part of quark distribution function.
Section 4 contains critical analysis of Vogt's arguments in
\cite{VogtinFreiburg}. In Section 5 the leading and
next--to--leading order QCD analysis of $F_2^{\gamma}$ is
formulated and the pointlike solutions of the inhomogeneous
evolution equation for $q_{\mathrm{NS}}^{\gamma}$ are
explicitly written down up to order $\alpha_s^2$. Phenomenological
implications of the present analysis for $F_2^{\gamma}(x,Q^2)$ are
outlined in Section 6, followed by the summary and conclusions in
Section 7.
Compared with \cite{factor} I have omitted the analysis of the
structure of the virtual photon, some of which can be found in
\cite{smarkem1,smarkem2}.

\section{Notation and basic facts}
In QCD the coupling of quarks and gluons is characterized by the
renormalized colour coupling (``couplant'' for short)
$\alpha_s(\mu)$, depending on the {\em renormalization scale} $\mu$
and satisfying the equation
\begin{equation}
\frac{{\mathrm d}\alpha_s(\mu)}{{\mathrm d}\ln \mu^2}\equiv
\beta(\alpha_s(\mu))=
-\frac{\beta_0}{4\pi}\alpha_s^2(\mu)-
\frac{\beta_1}{16\pi^2}
\alpha_s^3(\mu)+\cdots,
\label{RG}
\end{equation}
where, in QCD with $n_f$ massless quark flavours, the first two
coefficients, $\beta_0=11-2n_f/3$ and $\beta_1=102-38n_f/3$, are
unique, while all the higher order ones are ambiguous. As we shall
stay in this paper within the NLO QCD, only the first two, {\em
unique}, terms in (\ref{RG}) will be taken into account in the
following. However, even for a given r.h.s. of (\ref{RG}) its
solution $\alpha_s(\mu)$ is not a unique function of $\mu$,
because there is an infinite number of solutions of (\ref{RG}),
differing by the initial condition. This so called {\em
renormalization scheme} (RS) ambiguity \footnote{In higher orders
this ambiguity includes also the arbitrariness of the coefficients
$\beta_i,i\ge 2$.}
can be parameterized in a number of ways. One
of them makes use of the fact that in the process of
renormalization another dimensional parameter, denoted usually
$\Lambda$, inevitably appears in the theory. This parameter
depends on the RS and at the NLO even fully specifies it.
For instance, $\alpha_s(\mu)$ in
the familiar MS and $\overline{\mathrm {MS}}$ RS are two
solutions of the same equation (\ref{RG}), associated with different
$\Lambda_{\mathrm {RS}}$
\footnote{At the NLO the variation of both the
renormalization scale $\mu$ and the renormalization scheme
RS$\equiv$\{$\Lambda_{\mathrm {RS}}$\} is redundant. It
suffices to fix one of them and vary the other, but I stick
to the common practice of considering both of them as free
parameters.}. In this paper we shall work in the standard
$\overline{\mathrm {MS}}$ RS of the couplant.

``Dressed'' PDF
\footnote{In the following the adjective ``dressed'' as well as the
superscript ``$\gamma$'' will be dropped.}
result from the resummation of multiple parton collinear emission
off the corresponding ``bare'' parton distributions. As a result of
this resummation PDF acquire dependence on the {\em factorization
scale} $M$. This scale defines the upper limit on
some measure $t$ of the off--shellness of partons included in the
definition of $D(x,M)$
\begin{equation}
D_i(x,M)\equiv \int_{t_{\mathrm {min}}}^{M^2}{\mathrm d}t
d_i(x,t),~~~~~~i=q,\overline{q},G,
\label{dressed}
\end{equation}
where the {\em unintegrated} PDF $d_i(x,t)$ describe distribution
functions of partons with the momentum fraction $x$ and {\em fixed}
off--shellness $t$. Parton virtuality $\tau\equiv\mid p^2-m^2\mid$
or transverse mass $m_T^2\equiv p_T^2+m^2$, are two standard
choices of such a measure. Because at small $t$, $d_i(x,t)={\cal
O}(1/t^k), k=1,2$, the dominant part of the integral
(\ref{dressed}) comes from the region of small off--shellness $t$.
Varying the upper bound $M^2$ in (\ref{dressed}) has therefore only
a small effect on the integral (\ref{dressed}), leading to weak (at
most logarithmic) scaling violations.
Factorization scale dependence of PDF of the photon
\footnote{If not stated otherwise all distribution functions
in the following concern the photon.}
is determined by the system of coupled inhomogeneous evolution
equations
\begin{eqnarray}
\frac{{\mathrm d}\Sigma(x,M)}{{\mathrm d}\ln M^2}& =&
\delta_{\Sigma}k_q+P_{qq}\otimes \Sigma+ P_{qG}\otimes G,
\label{Sigmaevolution}
\\ \frac{{\mathrm d}G(x,M)}{{\mathrm d}\ln M^2} & =& k_G+
P_{Gq}\otimes \Sigma+ P_{GG}\otimes G, \label{Gevolution} \\
\frac{{\mathrm d}q_{\mathrm {NS}}(x,M)}{{\mathrm d}\ln M^2}& =&
\delta_{\mathrm {NS}} k_q+P_{\mathrm {NS}}\otimes q_{\mathrm{NS}},
\label{NSevolution}
\end{eqnarray}
where
\footnote{For nonsinglet quark
distribution function $q_{\mathrm{NS}}$ another definition is used
in the literature \cite{Aurenche}. The definition adopted here
corresponds to that used in \cite{GRVNS}.}
\begin{eqnarray}
\Sigma(x,M) & \equiv & \sum_{i=1}^{n_f}q_i^{+}(x,M)\equiv
\sum_{i=1}^{n_f} \left[q_i(x,M)+\overline{q}_i(x,M)\right],
\label{singlet}\\ q_{\mathrm{NS}}(x,M)& \equiv &
\sum_{i=1}^{n_f}\left(e_i^2-\langle e^2\rangle\right)
\left(q_i(x,M)+\overline{q}_i(x,M)\right),
\label{nonsinglet}
\end{eqnarray}
\begin{equation}
\delta_{\mathrm{NS}}=6n_f\left(\langle e^4\rangle-\langle
e^2\rangle ^2\right),~~~\delta_{\Sigma}=6n_f\langle e^2\rangle.
\label{sigmas}
\end{equation}
To order $\alpha$ the splitting functions $P_{ij}$ and
$k_i$ are given as power expansions in $\alpha_s(M)$:
\begin{eqnarray}
k_q(x,M) & = & \frac{\alpha}{2\pi}\left[k^{(0)}_q(x)+
\frac{\alpha_s(M)}{2\pi}k_q^{(1)}(x)+
\left(\frac{\alpha_s(M)}{2\pi}\right)^2k^{(2)}_q(x)+\cdots\right],
\label{splitquark} \\ k_G(x,M) & = &
\frac{\alpha}{2\pi}\left[~~~~~~~~~~~~
\frac{\alpha_s(M)}{2\pi}k_G^{(1)}(x)+
\left(\frac{\alpha_s(M)}{2\pi}\right)^2k^{(2)}_G(x)+\cdots\;\right],
\label{splitgluon} \\ P_{ij}(x,M) & = &
~~~~~~~~~~~~~~~~~~\frac{\alpha_s(M)}{2\pi}P^{(0)}_{ij}(x) +
\left(\frac{\alpha_s(M)}{2\pi}\right)^2 P_{ij}^{(1)}(x)+\cdots,
\label{splitpij}
\end{eqnarray}
where the leading order splitting functions
$k_q^{(0)}(x)=(x^2+(1-x)^2)$ and $P^{(0)}_{ij}(x)$ are {\em
unique}, while all higher order ones
$k^{(j)}_q,k^{(j)}_G,P^{(j)}_{kl},j\ge 1$ depend on the choice of
the {\em factorization scheme} (FS). The equations
(\ref{Sigmaevolution}-\ref{NSevolution}) can be recast into evolution
equations for $q_i(x,M),\overline{q}_i(x,M)$ and $G(x,M)$ with
inhomegenous splitting functions $k_{q_i}^{(0)}=3e_i^2k_q^{(0)}$.
The photon structure function $F_2^{\gamma}(x,Q^2)$, measured in
deep inelastic scattering of electrons on photons is given as
\begin{eqnarray}
\frac{1}{x}F_2^{\gamma}(x,Q^2)&=& q_{\mathrm{NS}}(M)\otimes
C_q(Q/M)+\frac{\alpha}{2\pi}\delta_{\mathrm{NS}}C_{\gamma}+
\label{NSpart} \\ & & \langle e^2\rangle \Sigma(M)\otimes
C_q(Q/M)+\frac{\alpha}{2\pi} \langle
e^2\rangle\delta_{\Sigma}C_{\gamma}+ \langle
e^2\rangle\frac{\alpha_s}{2\pi}G(M)\otimes C_G(Q/M)
\label{S+Gpart}
\end{eqnarray}
of photonic PDF and coefficient functions
$C_q(x),C_G(x),C_{\gamma}(x)$ admitting perturbative expansions
\begin{eqnarray}
C_q(x,Q/M) & = & \delta(1-x)~~~~+
~~~\frac{\alpha_s(\mu)}{2\pi}C^{(1)}_q(x, Q/M)+\cdots,
\label{cq} \\
C_G(x,Q/M) & = & ~~~~~~~~~~~~~~~~~~~~~
\frac{\alpha_s(\mu)}{2\pi}C^{(1)}_G(x,Q/M)+\cdots,
\label{cG} \\
C_{\gamma}(x,Q/M) & = &
C_{\gamma}^{(0)}(x,Q/M)+
\frac{\alpha_s(\mu)}{2\pi}C_{\gamma}^{(1)}(x,Q/M)+\cdots,
\label{cg}
\end{eqnarray}
where the standard formula for $C_{\gamma}^{(0)}$ reads
\footnote{Alternatives to this expression for $Q^2=M^2$
are discussed in Subsection 6.6.}
\begin{equation}
C_{\gamma}^{(0)}(x,Q/M)=\left(x^2+(1-x)^2\right)
\left[\ln\frac{M^2}{Q^2}+\ln\frac{1-x}{x}\right]+8x(1-x)-1.
\label{C0}
\end{equation}
The renormalization scale $\mu$, used as argument of
$\alpha_s(\mu)$ in (\ref{cq}-\ref{cg})
is in principle independent of the
factorization scale $M$. Note that despite the presence of $\mu$ as
argument of $\alpha_s(\mu)$ in (\ref{cq}--\ref{cg}), the
coefficient functions $C_q,C_G$ and $C_{\gamma}$ are actually
independent of $\mu$ because the $\mu$--dependence of the
expansion parameter
$\alpha_s(\mu)$ is cancelled by explicit dependence of $C^{(i)}_q,
C^{(i)}_G,C^{(i)}_{\gamma},i\ge 2$ on $\mu$ \cite{politzer}. On the
other hand, PDF and the coefficient functions $C_q, C_G$ and
$C_{\gamma}$ do depend on both the factorization scale $M$ and
factorization scheme, but in such a correlated manner that
physical quantities,
like $F_2^{\gamma}$, are independent of both $M$ and the FS,
provided expansions (\ref{splitquark}--\ref{splitpij}) and
(\ref{cq}--\ref{cg}) are taken to all orders in $\alpha_s(M)$ and
$\alpha_s(\mu)$. In practical calculations based on truncated forms
of (\ref{splitquark}--\ref{splitpij}) and (\ref{cq}--\ref{cg}) this
invariance is, however, lost and the choice of both $M$ and FS
makes numerical difference even for physical quantities. The
expressions for
$C_q^{(1)},C^{(1)}_G$ given in \cite{bardeen} are usually claimed
to correspond to ``$\overline{\mathrm {MS}}$ factorization scheme''.
As argued in \cite{jch2}, this denomination is, however,
incomplete. The adjective ``$\overline{\mathrm {MS}}$'' concerns
exclusively the choice of the RS of the couplant $\alpha_s$ and has
nothing to do with the choice of the splitting functions
$P^{(1)}_{ij}$. The choices of the
renormalization scheme of the couplant $\alpha_s$ and of the
factorization scheme of PDF are two completely independent
decisions, concerning two different and in general unrelated
redefinition procedures. Both are necessary in order to specify
uniquely the results of fixed order perturbative calculations, but
we may combine any choice of the RS of the couplant with any choice
of the FS of PDF. The coefficient functions
$C_q,C_G,C_{\gamma}$ depend on both of them,
whereas the splitting functions depend only on the latter. The
results given in \cite{bardeen} correspond to $\overline{\mathrm
{MS}}$ RS of the couplant but to the ``minimal subtraction'' FS of
PDF
\footnote{See Section 2.6 of \cite{FP}, in particular eq. (2.31),
for discussion of this point.}. It is therefore more appropriate to
call this full specification of the renormalization and factorization
schemes as ``$\overline{\mathrm {MS}}+{\mathrm {MS}}$ scheme''.
Although the phenomenological relevance of treating $\mu$ and $M$
as independent parameters has been demonstrated \cite{fontannaz},
I shall follow the usual practice and set $\mu=M$.

\section{Pointlike solutions and their properties}
The general solution of the evolution equations
(\ref{Sigmaevolution}-\ref{NSevolution}) can be written as the sum
of a particular solution of the full inhomogeneous equation and the
general solution of the corresponding homogeneous one, called
{\em hadronic} \footnote{Sometimes also called ``VDM part''
because it is usually modelled by PDF of vector mesons.} part. A
subset of the solutions of full evolution equations resulting from
the resummation of series of diagrams like those
in Fig. (\ref{figpl}), which
start with the pointlike purely QED vertex $\gamma\rightarrow
q\overline{q}$, are called {\em pointlike} (PL) solutions. In
writing down the expression for the resummation of diagrams in
Fig. \ref{figpl} there is a freedom in specifying some sort of
boundary condition. It is common to work within a subset of
pointlike solutions specified by the value of the scale $M_0$ at
which they vanish. In general, we can thus write
($D=q,\overline{q},G$)
\begin{equation}
D(x,M^2)= D^{\mathrm {PL}}(x,M^2)+D^{\mathrm{HAD}}(x,M^2).
\label{separation}
\end{equation}
Due to the fact that there is an infinite number of pointlike
solutions $q^{\mathrm{PL}}(x,M^2)$, the separation of quark and
gluon distribution functions into their pointlike and hadronic
parts is, however, ambiguous and therefore these concepts have
separately no physical meaning. In \cite{smarkem1} we
discussed numerical aspects of this ambiguity for the
Schuler--Sj\"{o}strand sets of parameterizations \cite{sas1}.

\FIGURE { \epsfig{file=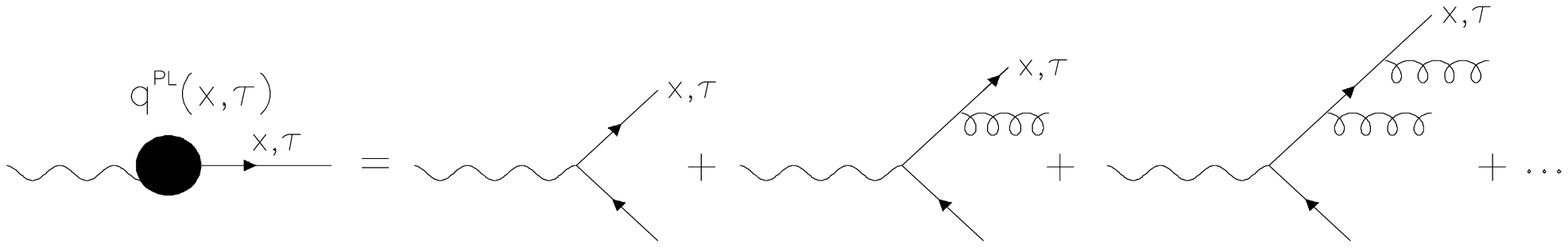,width=\textwidth}
\caption{Diagrams defining the pointlike parts of NS quark
distribution function in LL approximation. The resummation
involves integration over parton virtualities $M_0^2\le \tau\le
M^2$..} \label{figpl}}
\noindent
To see the most important feature of the
pointlike part of quark distribution functions that will be
crucial for the following analysis, let us consider in detail the
case of nonsinglet quark distribution function $q_{\mathrm
{NS}}(x,M)$, which is explicitly defined via the series
\begin{displaymath}
q^{\mathrm {PL}}_{\mathrm {NS}}(x,M_0,M) \equiv
\frac{\alpha}{2\pi}k_{\mathrm {NS}}^{(0)}(x)
\int^{M^2}_{M_0^2}\frac{{\mathrm d}\tau}{\tau}+
\int^{1}_{x}\frac{{\mathrm d}y}{y}P^{(0)}_{qq}
\left(\frac{x}{y}\right) \int^{M^2}_{M_0^2} \frac{{\mathrm
d}\tau_1}{\tau_1}\frac{\alpha_s(\tau_1)}{2\pi}\frac{\alpha}{2\pi}
k_{\mathrm {NS}}^{(0)}(y)\int^{\tau_1}_{M_0^2} \frac{{\mathrm
d}\tau_2}{\tau_2}+
\end{displaymath}
\begin{equation}
\int^{1}_{x}\frac{{\mathrm d}y}{y}P^{(0)}_{qq}
\left(\frac{x}{y}\right) \int^{1}_{y}\frac{{\mathrm
d}w}{w}P^{(0)}_{qq} \left(\frac{y}{w}\right) \int^{M^2}_{M_0^2}
\frac{{\mathrm d}\tau_1}{\tau_1}\frac{\alpha_s(\tau_1)}{2\pi}
\int^{\tau_1}_{M_0^2} \frac{{\mathrm
d}\tau_2}{\tau_2}\frac{\alpha_s(\tau_2)}{2\pi}\frac{\alpha}{2\pi}
k_{\mathrm {NS}}^{(0)}(w)\int^{\tau_2}_{M_0^2}\frac{{\mathrm
d}\tau_3}{\tau_3} +\cdots, \label{resummation}
\end{equation}
where $k_{\mathrm{NS}}^{(0)}(x)=\delta_{\mathrm{NS}}k^{(0)}_q(x)$.
In terms of moments defined as
\begin{equation}
f(n)\equiv \int_0^1x^n f(x)\mathrm{d}x
\label{moment}
\end{equation}
this series can be resummed in a closed form
\begin{equation}
q_{\mathrm {NS}}^{\mathrm {PL}}(n,M_0,M)=\frac{4\pi}{\alpha_s(M)}
\left[1-\left(\frac{\alpha_s(M)}{\alpha_s(M_0)}\right)^
{1-2P^{(0)}_{qq}(n)/\beta_0}\right]a_{\mathrm {NS}}(n),
\label{generalpointlike}
\end{equation}
where
\begin{equation}
a_{\mathrm {NS}}(n)\equiv \frac{\alpha}{2\pi\beta_0}
\frac{k_{\mathrm {NS}}^{(0)}(n)}{1-2P^{(0)}_{qq}(n)/\beta_0}.
\label{ans}
\end{equation}
\FIGURE{ \epsfig{file=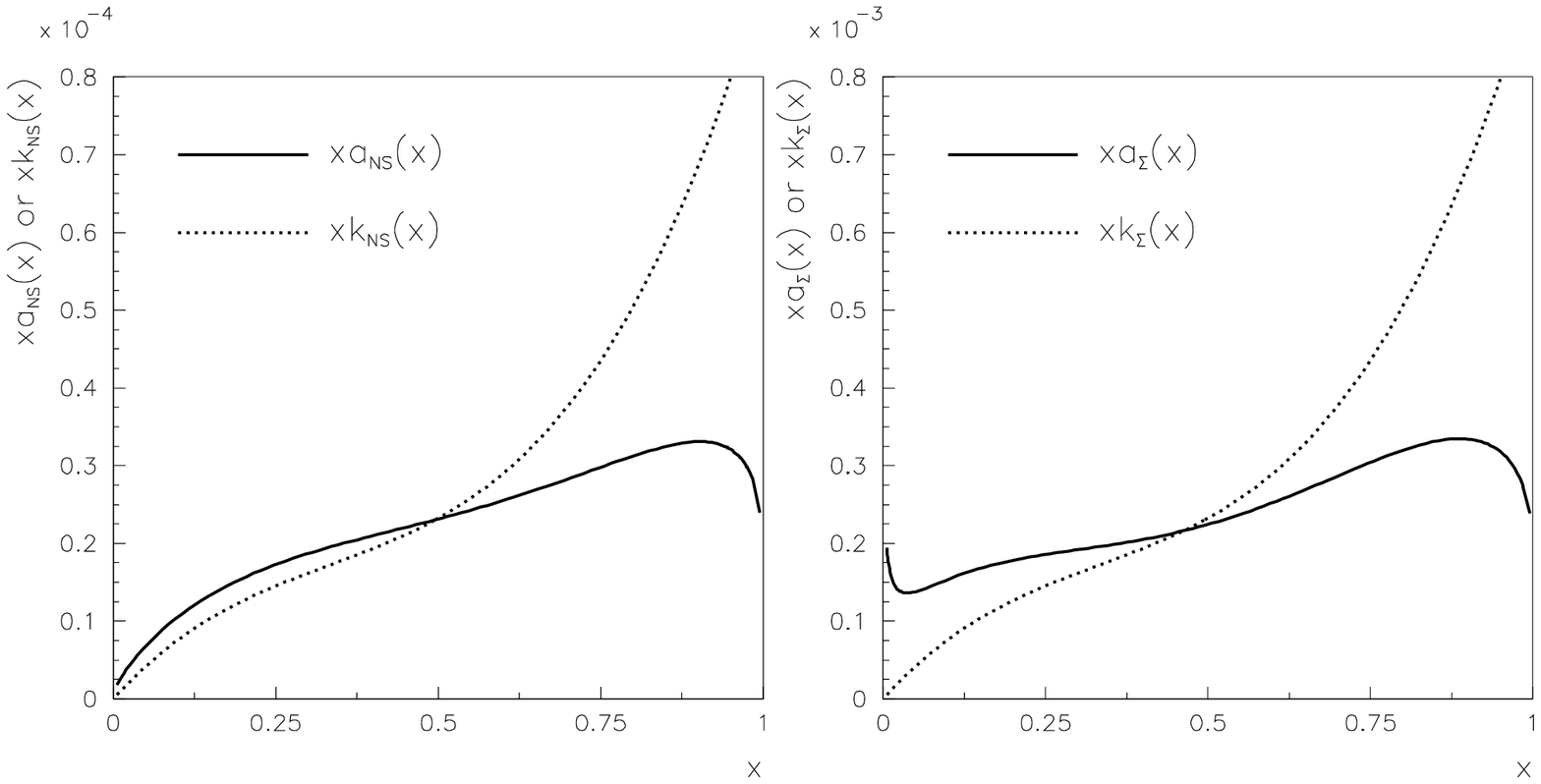,width=\textwidth}
\caption{Comparison, left in the nonsinglet and right in the
singlet channels, of the functions
$xk_i(x)=x\delta_ik_q^{(0)},i=\mathrm{NS},\Sigma$ with the
functions $xa(x)$ corresponding to asymptotic pointlike solutions
(\ref{asymptotic}) and its analogue in the singlet channel.}
\label{anskns}} It is straightforward to show that
(\ref{resummation}) or, equivalently, (\ref{generalpointlike}),
satisfy the evolution equation (\ref{NSevolution}) with the
splitting functions $k_q$ and $P_{ij}$ including the first terms
$k_q^{(0)}$ and $P_{qq}^{(0)}$ only.

Transforming (\ref{ans}) to the $x$--space by means of inverse Mellin
transformation we get
$a_{\mathrm{NS}}(x)$ shown in Fig. 2. The resummation softens
the $x-$dependence of $a_{\mathrm {NS}}(x)$ with
respect to the first term in (\ref{resummation}), proportional to
$k_{\mathrm {NS}}(x)$, but does not change the logarithmic
dependence of $q_{\mathrm {NS}}$ on $M$. In the nonsinglet
channel the effects of gluon radiation on
$q_{\mathrm{NS}}^{\mathrm{PL}}$
are significant for $x>0.6$ but small elsewhere, whereas
in the singlet
channel such effects are marked also for $x<0.5$,
where they lead to a steep rise of $xq_{\mathrm{NS}}^{\mathrm{PL}}$
at very small $x$. As emphasized long time
ago by authors of \cite{FKP} the logarithmic dependence of
$q_{\mathrm{NS}}^{\mathrm{PL}}$ on $\ln M$ has
nothing to do with QCD and results exclusively from integration
over the transverse momenta (virtualities) of quarks coming from
the basic QED $\gamma\rightarrow q\overline{q}$ splitting.
For $M/M_0\rightarrow \infty$ the second term in brackets of
(\ref{generalpointlike}) vanishes and therefore all pointlike
solutions share the same large $M$ behaviour
\begin{equation}
q^{\mathrm {PL}}_{\mathrm {NS}}(x,M_0,M)
\rightarrow \frac{4\pi}{\alpha_s(M)}a_{\mathrm {NS}}(x)\equiv
q^{\mathrm {AP}}_{\mathrm {NS}}(x,M)\propto \ln \frac{M^2}{\Lambda^2},
\label{asymptotic}
\end{equation}
defining the so called {\em asymptotic pointlike} solution
$q^{\mathrm {AP}}_{\mathrm {NS}}(x,M)$ \cite{witten,BB}. Note that
(\ref{asymptotic}) can be interpreted as a
special case of the general pointlike solution
(\ref{generalpointlike})
in which the lower integration limit $M_0$ in (\ref{resummation})
is set equal to $\Lambda$, i.e. $M_0=\Lambda$! The fact that for
the asymptotic pointlike solution (\ref{asymptotic}) $\alpha_s(M)$
appears in the denominator of (\ref{asymptotic}) has been the
source of misleading claims (see, for instance, \cite{Vogt}) that
$q(x,M)={\cal O}(1/\alpha_s)$. This claim is wrong for a number of
reasons. First, it is manifestly invalid for the widely used
Schuler--Sj\"{o}strand SaS1 and SaS2 sets, which take $M_0=0.6$
GeV and $M_0=2$ GeV. It is obvious \cite{jch1} that provided $M_0$
is kept fixed when $\alpha_s\rightarrow 0$ the sum
(\ref{resummation}) approaches its first term, i.e.
\begin{equation}
 q^{\mathrm {PL}}_{\mathrm {NS}}(x,M,M_0)\rightarrow
\frac{\alpha}{2\pi} k_{\mathrm {NS}}^{(0)}(x)\ln\frac{M^2}{M_0^2},
\label{QEDlimit}
\end{equation}
corresponding to purely QED splitting $\gamma\rightarrow
q\overline{q}$. However, the claim that $q\propto 1/\alpha_s$ is
misleading even for the asymptotic pointlike solution
(\ref{asymptotic}). The fact that it diverges when
$\Lambda\rightarrow 0$ is a direct consequence of the fact that
for this (and only this) pointlike solution the decrease of the
coupling $\alpha_s(M/\Lambda)$ as $\Lambda\rightarrow 0$ is
compensated by the simultaneous extension of the integration
region as $M_0=\Lambda\rightarrow0$! If, however, QCD is switched
off by sending $\Lambda\rightarrow 0$ {\em without} simultaneously
extending the integration region, i.e. for {\em fixed $M_0$},
there is no trace of QCD left and we get back the simple QED
formula (\ref{QEDlimit}). This suggests to identify QCD
contribution to NS quark distribution function with the difference
\begin{equation}
q^{\mathrm{QCD}}_{\mathrm{NS}}(x,M)\equiv
q_{\mathrm{NS}}^{\mathrm{PL}}(x,M_0,M)- \frac{\alpha}{2\pi}
k_{\mathrm {NS}}^{(0)}(x)\ln\frac{M^2}{M_0^2}. \label{difference}
\end{equation}
Expanding (\ref{generalpointlike}) in powers of $\alpha_s(M)$,
keeping $M_0$ and $M$ fixed we find
\begin{equation}
q^{\mathrm{QCD}}_{\mathrm{NS}}(n,M)=
\frac{\alpha}{2\pi}\frac{\alpha_s}{2\pi}\frac{k_q^{(0)}(n)
P_{qq}^{(0)}(n)}{2}\ln^2\frac{M^2}{M_0^2}+O(\alpha_s^2),
\label{smallalphas}
\end{equation}
as expected from the explicit expression for the second term on
the r.h.s. of (\ref{resummation}). Note that the leading term in
(\ref{smallalphas}) is proportional to the second power of
$L_M\equiv \ln(M^2/M_0^2)$. In subsection 5.2.2 I will argue that
in a complete LO QCD analysis $q_{\mathrm{NS}}$
contains another term that behaves as
$\alpha_s L_M$.

In summary, $q^{\mathrm{PL}}_{\mathrm {NS}}(x,M)$
can be written a sum of two terms: purely QED
contribution, proportional to $\alpha$, and the term describing
genuine QCD effects, proportional to $\alpha\alpha_s$ and thus
vanishing when QCD is switched off. One can use the usual claim
that $q\propto \alpha/\alpha_s$ merely as a shorthand for the
specification of large $M$ behaviour of (\ref{resummation}), as
expressed in (\ref{asymptotic}). In fact, this is in fact
what one finds
in the original papers \cite{witten,BB}\footnote{See, for
instance, eq. (3.20) of \cite{BB}.} which do not contain any
explicit claim that $q(x,M)={\cal O}(1/\alpha_s)$.

\section{What is proving Vogt's proof?}
Before presenting my reformulation of the LO and NLO QCD
analysis of
$F_2^{\gamma}$, let me go through Vogt's arguments
\footnote{Note that due to slightly different notation my
$C_{\gamma}^{(0)}$ plays the role of $C_{\gamma}^{(1)}$ in
\cite{VogtinFreiburg}.}
that purport to prove that $q(M)\propto \alpha/\alpha_s$.
For the purpose of the discussion in this Section I will adopt
Vogt's simplified notation, in which $a_s\equiv\alpha_s/4\pi$,
$F^{\gamma}\equiv
F_{2,{\mathrm{NS}}}^{\gamma}/\alpha$, $q^{\gamma}\equiv
q_{\mathrm{NS}}^{\gamma}$ and
${\mathrm{d}}F^{\gamma}/{\mathrm{d}}\ln Q^2=\dot{F}^{\gamma}$,
all obvious indices as well as overall charge factors are suppressed
and products of $x$--dependent quantities are understood as
convolutions in $x$ space or simple products in momentum space.
We have
\begin{eqnarray}
F^{\gamma}& = & C_q q^{\gamma}+C_{\gamma} \label{Fgamma}\\
\dot{q}^{\gamma} & = & k+Pq^{\gamma} \label{qdot}\\ \dot{F}^{\gamma}
& = & \dot{C}_q
q^{\gamma}+C_q\dot{q}^{\gamma}+\dot{C}_{\gamma}\label{Fdot}
\end{eqnarray}
The r.h.s. of (\ref{Fdot}) can be rewritten as \cite{VogtinFreiburg}
\begin{equation}
\dot{F}^{\gamma}=\left(-\dot{C}_qC_q^{-1}C_{\gamma}+
C_qk-PC_{\gamma}+\dot{C}_{\gamma}\right)
+\left(\dot{C}_qC_q^{-1}+P\right)F^{\gamma},\label{second}
\end{equation}
where expressions in the brackets are factorization scheme
invariants. Inserting into (\ref{second}) perturbation expansions of
$C_q, C_{\gamma}$, their derivatives and the splitting functions
$k,~P$ we get
\footnote{
Note that eq. (14) of \cite{VogtinFreiburg} contains two misprints
in the bracket standing by $a_s^2$: the last term comes
with the wrong sign and the preceding one should correctly read
$-\beta_0C_{\gamma}^{(2)}$ (my $C_{\gamma}^{(1)}$ and
$C_{\gamma}^{(0)}$ correspond to $C_{\gamma}^{(2)}$ and
$C_{\gamma}^{(1)}$ in Vogt's notation).}
\newpage
\begin{eqnarray}
\lefteqn{
\dot{F}^{\gamma} = ~~(a_s)^0k^{(0)}} & & \label{a0}\nonumber\\
 & &
 +(a_s)^1\left[k^{(1)}+C_q^{(1)}k^{(0)}-P^{(0)}C_{\gamma}^{(0)}
 \right]
 \label{a1}\\
 & &
 +(a_s)^2\left[k^{(2)}+C_q^{(1)}k^{(1)}+
 C_q^{(2)}k^{(0)}-P^{(0)}C_{\gamma}^{(1)}-P^{(1)}C_{\gamma}^{(0)}
 -\beta_0C_{\gamma}^{(1)}
 +\beta_0C^{(1)}_qC^{(0)}_{\gamma}\right]\label{a2}
 \nonumber\\
  & & +F^{\gamma}[a_sP^{(0)}+\cdots].\label{af}\nonumber
\end{eqnarray}
The fact that
$C_{\gamma}^{(0)}$ appears in the expression by $a_s$
in (\ref{a1}) together with $C_q^{(1)}k^{(0)}$, in Vogt's words
that ``$C_{\gamma}^{(0)}$ .... enters $\dot{F}^{\gamma}$ on the
same level as the hadronic NLO quantity $C_q^{(1)}$'' leads him to
the conclusion that
``$C_{\gamma}^{(0)}$ is to be considered as a NLO contribution'',
and, consequently, `` for the purpose of power counting in $\alpha_s$
 the quark densities $q^{\gamma}$ and the structure function
$F_2^{\gamma}$ have to counted as $1/\alpha_s$.''
The flaw of this argument is obvious if one applies it to
$C_{\gamma}^{(1)}$ ($C_{\gamma}^{(2)}$ in Vogt's notation) appearing
in the expression standing in (\ref{a1}) by $a_s^2$ accompanied by
both $C_q^{(2)}$ and $C_q^{(1)}$. Repeating Vogt's argument leads
us to contradictory
conclusions that $C_{\gamma}^{(1)}$ is simultaneously of the same
order as $C_q^{(1)}$ and $C_q^{(2)}$! The resolution of this
contradiction is obvious: we have to take into account
the fact that $C_{\gamma}^{(0)}$ and $C_{\gamma}^{(1)}$ do not enter
the coefficients in (\ref{a1}) alone, but in products with other
quantities, corresponding to different orders of perturbative QCD.
For instance, since $k^{(0)}$ stands in (\ref{splitquark})
by $(\alpha_s)^0$ whereas $k^{(1)}$ by $\alpha_s$, the products
$C_q^{(1)}k^{(1)}$ and $C_q^{(2)}k^{(0)}$ are of the same order
\footnote{Which is one order of $\alpha_s$ higher than
$C_q^{(1)}k_q^{(0)}$.},
despite the fact that $C_q^{(1)}$ and $C_q^{(2)}$ stand in
(\ref{cq}) by $\alpha_s$ and $\alpha_s^2$ respectively. Similarly,
$C_{\gamma}^{(1)}$ is, as expected, of one order of $\alpha_s$
higher than $C_{\gamma}^{(0)}$. The failure of Vogt to take this
fact into account led him to wrong conclusion concerning
$C_{\gamma}^{(0)}$. Taking into account that
$k^{(0)}$ stands by $(\alpha_s)^0$ whereas $P^{(0)}$ by $\alpha_s$
implies that $C_{q}^{(1)}$ is of one order of $\alpha_s$ higher than
$C_{\gamma}^{(0)}$ and thus $C_{\gamma}^{(0)}$ not of the NLO!

If I am right, why have most theorists
\footnote{With exception  of the authors of \cite{FKP}, who have
advocated ideas closely related to those advanced in this paper, for
a long time.}
so tenaciously clung to the claim that in some sense
$q\propto 1/\alpha_s$? In part because it provides a simple way of
expressing large scale behaviour of $q(x,M)$, but the main reason
is related to another tenet of the conventional
approach to $F_2^{\gamma}(x,Q^2)$, namely the assumption that at the
leading order of QCD the observable $F_2^{\gamma}(x,Q^2)$ is related
to $q(x,M)$ in the same way as for hadrons, i.e.
\begin{equation}
F_{2,\mathrm{LO}}^{\gamma}(x,Q^2)=q_{\mathrm{LO}}(x,M^2).
\label{F2wrong}\
\end{equation}
This -- incorrect as I am just going to argue --
relation is in fact the
principal source of all misleading and factually wrong statements
concerning the QCD analysis of $F_2^{\gamma}$.

To evaluate the r.h.s. of (\ref{F2wrong}) the factorization scale $M$
has to be chosen. The requirement of factorization scale invariance
of physical quantities implies that at any finite order of
perturbative QCD the difference $PQ^{(n)}(M_1)-PQ^{(n)}(M_2)$ of
perturbative predictions for a physical quantity $PQ$ evaluated up
to order $\alpha_s^n$ at two scales $M_1$ and $M_2$ behaves as
$O(\alpha_s^{n+1})$. For hadrons (\ref{F2wrong}) is consistent
with this fundamental requirement due to the fact that for them
$q(x,M)$ satisfies homogeneous evolution equations.
Consequently, the difference
\begin{equation}
\Delta(M_1,M_2)\equiv q(M_1)-q(M_2)\propto \alpha_s(M_1)
\ln \left(\frac{M_1^2}{M_2^2}\right)q(M_1)
\label{consistency}
\end{equation}
is, indeed, of one order of $\alpha_s$ higher that $q$ itself and
therefore (\ref{F2wrong})
to the order considered independent of the choice of $M$!

For the photon the presence of the
inhomogeneous term $(\alpha/2\pi)k_q^{(0)}$ in the evolution
equations for quark distribution functions implies, however,
$\Delta(M_1,M_2)\propto \alpha\ln (M_1^2/M_2^2)$ and, consequently,
$F_{2,\mathrm{LO}}^{\gamma}(M_1)-F_{2,\mathrm{LO}}^{\gamma}(M_1)
\propto \alpha$. To retain the factorization scale invariance of
(\ref{F2wrong}) for $F_{2,\mathrm{LO}}^{\gamma}$ the property
$q\propto 1/\alpha_s$ seems therefore indispensable! The use of
the term ``NLO'' for $C_{\gamma}^{(0)}$
is also sometimes justified by the fact that for fixed ratio
$d\equiv M_1/M_2$, the $M^2$ dependence of difference
$\Delta(M_1,M_2)\propto \alpha \ln d$ is ``subleading'' with respect
to $q_{\mathrm{LO}}\propto \alpha\ln M^2$. This is true but it must
be kept in mind that the dominant $\ln M^2$ behaviour of the standard
definition of $q_{\mathrm{LO}}$ is basically a consequence of QED,
not QCD dynamics! To claim with Vogt that the purely QED term
$C_{\gamma}^{(0)}(Q^2=M^2)$ is of the ``NLO'' merely because it is
$M^2$--independent constant and thus small for large $M^2$ with respect
to $q_{\mathrm{QED}}\propto \alpha \ln M^2$
is a misuse of the terms ``LO'' and ``NLO'',
which are meant to describe different orders in $\alpha_s$, not
different large $M^2$ behaviours.

In the conventional approach to $F_2^{\gamma}$ wrong medicine is thus
used to salvage the factorization scale invariance
of (\ref{F2wrong}). There is, however, a different and consistent way
of guaranteeing this invariance at the LO that
does not rely on the untenable assumption $q\propto 1/\alpha_s$:
to modify the relation (\ref{F2wrong}) itself!
The rest of this paper is devoted to the elaboration of this
claim. I will show that the modified relation between
$F_{2,\mathrm{LO}}^{\gamma}$ and $q_{\mathrm{LO}}$ is consistent with
the obvious fact that $q\propto \alpha$ and demonstrate how the
additional terms in this relation conspire to quarantee
the factorization scale invariance of $F^{\gamma}_{2,\mathrm{LO}}$.

\section{QCD analysis of ${\mathbf{F_2^{\gamma}(x,Q^2)}}$}
In this Section QCD analysis of $F_2^{\gamma}(x,Q^2)$
that separates genuine QCD effects from those of purely QED origin
will be presented. Throughout this and following sections I will
restrict my attention to the nonsiglet part (\ref{NSpart}) of
$F_2^{\gamma}$
\begin{equation}
\frac{1}{x}F_{2,\mathrm{NS}}^{\gamma}(x,Q^2)= q_{\mathrm{NS}}(M)
\otimes C_q(Q/M)+\frac{\alpha}{2\pi}\delta_{\mathrm{NS}}C_{\gamma}
\label{F2NS}
\end{equation}
and nonsinglet quark distribution function as defined in
(\ref{nonsinglet}). I will consider the contribution of
light quarks (i.e. $u,d,s$) only and, moreover, disregard the
difference
\footnote{This difference is entirely negligible above roughly
$x=0.25$ but becomes sizable below this value.}
between their distributions functions after division
by $e_q^2$. Under these simplifying assumptions we have
\begin{eqnarray}
\lefteqn{
\frac{1}{x}F_{2,\mathrm{NS}}^{\gamma}(x,Q^2)=\delta_{\mathrm{NS}}
\left[q(M)\otimes C_q(Q/M)+\frac{\alpha}{2\pi}
C_{\gamma}(Q/M)\right]}
 & & \label{F2NSsimple}\\
& = & \delta_{\mathrm{NS}}\left[q(M)+\frac{\alpha_s}{2\pi}q(M)
\otimes C_q^{(1)}(Q/M)+\frac{\alpha}{2\pi}C_{\gamma}^{(0)}(Q/M)+
\frac{\alpha}{2\pi}\frac{\alpha_s}{2\pi}C_{\gamma}^{(1)}(Q/M)\cdots
\right]
\label{jks}\nonumber
\end{eqnarray}
where $q\equiv u(x)/3e_u^2\doteq d(x)/3e_d^2\doteq s(x)/3e_s^2$.
For simplicity I will in the following drop the overall charge factor
$\delta_{\mathrm{NS}}$ as well as the subscript ``NS'' and
continue to use the dot to denote the derivatives with respect
to $\ln M^2$. In (\ref{F2NS}--\ref{F2NSsimple}) I have written out
explicitly the symbol $\otimes$ to denote convolutions in $x$-space,
but henceforth I will work mostly in the momentum space and thus all
products of quark distribution and coefficient functions are
understood as simple multiplications in momentum space
\footnote{To save space, their dependence on the momentum variable $n$
will not be written out explicitly.}.

\subsection{Defining leading and
next--to--leading orders for ${\mathbf{F_2^{\gamma}(x,Q^2)}}$}
Let us first clearly define what is meant under the terms ``leading''
and ``next--to--leading'' orders in the case of QCD analysis of
$F_2^{\gamma}$. As for hadronic parts of photonic PDF
the definition of these terms is the same as for hadrons, I will
concentrate in this Section on the properties of the pointlike
quark distribution function and its contribution to $F_2^{\gamma}$.

It is useful to recall the meaning of the terms ``leading'' and
``next--to--leading'' order of perturbative QCD for
the case of the familiar ratio
\begin{equation}
R_{\mathrm{e}^+\mathrm{e}^-}(Q)\equiv
\frac{\sigma(\mathrm{e}^+\mathrm{e}^-\rightarrow
\mathrm{hadrons})}{\sigma(\mathrm{e}^+\mathrm{e}^-\rightarrow
\mu^+\mu^-)}
=\left(3\sum_{i=1}^{n_f}e_i^2\right)\left(1+r(Q)\right)
\label{Rlarge}
\end{equation}
measured in e$^+$e$^-$ annihilations.
The prefactor
\begin{equation}
R_{\mathrm{QED}}\equiv \left(3\sum_{i=1}^{n_f}e_i^2\right)
\label{RQED}
\end{equation}
comes from
pure QED, whereas genuine QCD effects give $r(Q)$ as perturbation
expansion in powers of $\alpha_s$
\begin{equation}
r(Q)=\frac{\alpha_s(M)}{\pi}\left[1+\frac{\alpha_s(M)}
{2\pi}r_1(Q/M)+\cdots\right].
\label{rsmall}
\end{equation}
For the quantity (\ref{Rlarge}) it is a generally accepted procedure
to divide out the QED contribution $R_{\mathrm{QED}}$
and apply the terms ``leading'' and ``next--to--leading'' only to
QCD analysis of $r(Q)$, which starts as $\alpha_s/\pi$.
Nobody would suggest including $R_{\mathrm{QED}}$ in the definition
of the term ``leading'' order in QCD analysis of (\ref{Rlarge}).

Unfortunately, precisely this is conventionally done in the case
of $F_2^{\gamma}$. I will now present the organization of QCD
expression for $F_2^{\gamma}$ that follows as closely as possible
the convention used in QCD analysis of (\ref{Rlarge}). For this
purpose let us write, following the discussion at the end of
Section 3, the pointlike quark distribution function $q(M)$,
satisfying the condition $q(M_0)=0$, as a sum of two terms
\begin{equation}
q(M)=q_{\mathrm{QED}}(M)+q_{\mathrm{QCD}}(M)
\label{qQEDQCD}
\end{equation}
where the purely QED contribution is given as
\begin{equation}
q_{\mathrm{QED}}(M)\equiv\frac{\alpha}{2\pi}k_q^{(0)}
\ln\frac{M^2}{M_0^2}. \label{qQED}
\end{equation}
$M_0$, a free parameter specifying the pointlike solution, can be
interpreted as the lower limit on the integral over quark
virtuality included in the definition of $q_{\mathrm{QED}}(M)$.
Defined in this way $q_{\mathrm{QCD}}$ is due entirely to QCD
effects, i.e. vanishes when we switch off QCD.
The expression (\ref{qQED}) is a close analogue of the
QED contribution $R_{\mathrm{QED}}$ in (\ref{RQED}). Note that
$q_{\mathrm{QCD}}$ satisfies the inhomogeneous evolution equation
\begin{eqnarray}
\dot{q}_{\mathrm{QCD}}& = &
\frac{\alpha_s}{2\pi}\left[\frac{\alpha}{2\pi}k^{(1)}_q+P^{(0)}_{qq}
q_{\mathrm{QED}}\right]+
\left(\frac{\alpha_s}{2\pi}\right)^2\left[
\frac{\alpha}{2\pi}k_q^{(2)}+P_{qq}^{(1)}q_{\mathrm{QED}}\right]
\cdots + \label{inh}\nonumber\\
& & \frac{\alpha_s}{2\pi}P^{(0)}_{qq}q_{\mathrm{QCD}}+
\left(\frac{\alpha_s}{2\pi}\right)^2P^{(1)}_{qq}q_{\mathrm{QCD}}+\cdots,
\label{modified}
\end{eqnarray}
which differs from that satisfied by the full quark distribution
function not only by the absence of the first term
$(\alpha/2\pi)k_q^{(0)}$ but also by shifted appearance of higher
order coefficients $k^{(i)}_q;i\ge 1$. For instance, the
inhomogeneous splitting function $k^{(1)}_q$ enters the evolution
equation for $q_{\mathrm{QCD}}$ at the same order as homogeneous
splitting function $P_{qq}^{(0)}$ and thus these splitting
functions will appear at the same order also in its solutions.
The fact that $k_q^{(1)}$ is
a function of $x$ (or $n$) only, whereas $q_{\mathrm{QED}}$ is in
addition also a function of $M$, influences relative importance of
the two terms making up the coefficient by $\alpha_s$ in
(\ref{modified}) as $M$ varies, but does not change the basic
observation, namely that both of them contribute at the same
order. Similar statement holds for all pairs
$k_q^{(i+1)},P^{(i)}_{qq};i\ge 0$. The simultaneous presence of
$k^{(2)}_q$ and $P^{(1)}_{qq}$ in the $O(\alpha_s^2)$ term of the
inhomogeneous part of (\ref{modified}) is yet another
expression of my claim in \cite{factor} that NLO QCD analysis
of $F_2^{\gamma}$ requires the knowledge of $k^{(2)}_q$. I have,
however, failed to realize that the same argument implies that in
a complete LO QCD analysis the evolution equation for quark
distribution function must include the $O(\alpha_s)$ inhomogeneous
splitting function $k_q^{(1)}$ as well. Explicit expressions of
the resulting solutions are presented and their properties
discussed below.

\subsubsection{QED part}
The purely QED result
\footnote{ In which the integration over virtual quark
virtualities is restricted to $\tau\ge M_0^2$.} for
$F_2^{\gamma}$ is given by the sum
\begin{equation}
\frac{1}{x}F_{2,\mathrm{QED}}^{\gamma}(x,Q^2)=q_{\mathrm{QED}}(x,M)+
\frac{\alpha}{2\pi}C_{\gamma}^{(0)}(x,Q/M),
\label{FQED}
\end{equation}
which is a close analogue of QED prediction (\ref{RQED}) for the
ratio (\ref{Rlarge}). Note that although both terms on the r.h.s.
of (\ref{FQED}) depend on factorization scale $M$, their sum is
independent of it and is a function of $Q^2$ and $M_0^2$
only! \footnote{The dependence on $M_0$ must not be confused with
factorization scale dependence.}

Let me emphasize that already at this point I depart from the
conventional approach which identifies $C_{\gamma}^{(0)}$ as part
of the ``NLO'' corrections. In fact $C_{\gamma}^{(0)}$ has nothing
to do with QCD at all and is entirely due to QED effects! It must
therefore be always present in any QCD analysis of data.
It is thus also not true, as claimed in \cite{marco},
that the SaS1M and SaS2M parameterizations
\footnote{In fact the SaS1M and SaS2M parameterizations
include in $F_2^{\gamma}$ only part of the expression (\ref{C0}).
This and related facts are discussed in detail in Subsection
6.6.}
are ``theoretically inconsistent'' because they combine in LO
expression for $F_{2}^{\gamma}$ the ``NLO'' quantity
$C_{\gamma}^{(0)}$ with the LO quark distribution functions.
In fact, just the opposite is true! From the point of view of
including $C_{\gamma}^{(0)}$ the analysis of $F_2^{\gamma}$ in
terms of SaS1M and SaS2M parameterizations are closer to complete
LO QCD analysis than analogous analysis using SaS1D or SaS2D
sets.

Let me reiterate that the principal feature of
$F_2^{\gamma}(x,Q^2)$, namely its logarithmic rise with $Q^2$,
is basically a QED effect and consequently its observation in
experiments by itself no check of QCD. One has to go to subtler
features \footnote{For instance, the deviation of the
$x$--dependence of
$\mathrm{d}F_2^{\gamma}(x,Q^2)/\mathrm{d}\ln Q^2$ from the QED
result $(\alpha/2\pi)k^{(0)}_q(x)$.} to identify genuine QCD
effects!

\subsubsection{QCD part}
In conventional analyses of $F_2^{\gamma}$ the first term on the
r.h.s. of (\ref{FQED}) is included in the ``leading'' order
of QCD thereby
obscuring meaning of the term ``leading''. There is no obstacle to
following the procedure adopted in QCD analyses of (\ref{Rlarge})
and rewrite also $F_2^{\gamma}(x,Q^2)$ as the sum of its QED and
QCD parts
\begin{eqnarray}
\frac{1}{x}F_2^{\gamma}&=&
\underbrace{q_{\mathrm{QED}}+\frac{\alpha}{2\pi}C_{\gamma}^{(0)}}
_{\mathrm{pure~QED}}+
\label{F2QED}\\
& &
\underbrace{
q_{\mathrm{QCD}}+\frac{\alpha_s}{2\pi}C_q^{(1)}q_{\mathrm{QED}}
+\frac{\alpha}{2\pi}\frac{\alpha_s}{2\pi}C_{\gamma}^{(1)}}_
{\equiv A_1,~\mathrm{starting~as}~O(\alpha\alpha_s)}+\label{F2QCDLO}\\
 & &
\underbrace{
\frac{\alpha_s}{2\pi}C_q^{(1)}q_{\mathrm{QCD}}+\frac{\alpha}{2\pi}
\left(\frac{\alpha_s}{2\pi}\right)^2C_{\gamma}^{(2)}+
\left(\frac{\alpha_s}{2\pi}\right)^2C_q^{(2)}q_{\mathrm{QED}}
}_{\equiv A_2,~\mathrm{starting~ as}~ O(\alpha\alpha_s^2)}
\label{F2QCDNLO}
\end{eqnarray}
In (\ref{F2QED}-\ref{F2QCDNLO}) I have grouped into quantities
$A_0,~A_1$ and $A_2$ the contributions that start at zero, first
and second order of $\alpha_s$, assuming
$q_{\mathrm{QCD}}\approx \alpha_s$. In the next subsection I will
present explicit solutions demonstrating this behaviour.

The first line in (\ref{F2QED}) contains the purely QED contribution
(\ref{F2QED}) to $F_2^{\gamma}$. Following the analogy with
(\ref{Rlarge}) the LO QCD approximation to $F_2^{\gamma}$ is
identified with $A_1$,
\begin{equation}
\frac{1}{x}F_{2,\mathrm{LO}}^{\gamma}(x,Q^2)=
q_{\mathrm{QCD}}+\frac{\alpha_s}{2\pi}C_q^{(1)}q_{\mathrm{QED}}
+\frac{\alpha}{2\pi}\frac{\alpha_s}{2\pi}C_{\gamma}^{(1)},
\label{F2LO}
\end{equation}
the NLO one with the sum $A_1+A_2$, etc..

The expression (\ref{F2LO}) represents an analogue of the LO
contribution to (\ref{rsmall}), equal to $r_{\mathrm{LO}}=
\alpha_s/\pi$. Beside
its more complicated structure eq. (\ref{F2LO}) differs from
$r_{\mathrm{LO}}$ also by the fact that whereas the latter is
unique, (\ref{F2LO}) depends on $M_0$. This means that any QCD
analysis of $F_2^{\gamma}$ must start with fixing the value of
this parameter. It is worth emphasizing that for heavy
quarks $M_0^2$ can be related to  $m_Q^2$ and for
virtual photons to their virtuality $P^2$.

As my criticism in Section 4 of the conventional formulation of
QCD analysis of $F_2^{\gamma}$ was based on the lack of the
factorization scale invariance of the latter, let us now check
whether this invariance holds for the expression (\ref{F2LO})
where $q_{\mathrm{QCD}}$ satisfies (\ref{modified}) up to
order $\alpha_s$. Taking the derivative of
(\ref{F2LO}) with respect to $M^2$ we find, taking into account
that from similar considerations in hadronic collisions we know
that $\dot{C}_q^{(1)}=-P_{qq}^{(0)}$,
\begin{equation}
\dot{F}^{\gamma}_{\mathrm{LO}}=
\frac{\alpha}{2\pi}\frac{\alpha_s}{2\pi}\left[k_q^{(1)}+C_q^{(1)}+
\dot{C}^{(1)}_{\gamma}\right]+ O(\alpha_s^2). \label{dF2LO}
\end{equation}
Contrary to the case of hadronic structure function
$F_2^{\mathrm{p}}$, where analogous derivative is proportional to
$\alpha_s F_2^{\mathrm{p}}$ and thus manifestly of NLO, for
(\ref{F2LO}) the condition that (\ref{dF2LO}) is of order
$\alpha_s^2$ implies the following nontrivial relation
\begin{equation}
k_q^{(1)}+C_q^{(1)}+ \dot{C}^{(1)}_{\gamma}=0
\label{condition}
\end{equation}
between the quantities $k_q^{(0)}, k_q^{(1)}$ and
$C_{\gamma}^{(1)}$. Because $C_{\gamma}^{(1)}$ depends on $\ln
M^2$ as $\ln^2(Q^2/M^2)$ (\ref{condition}) actually implies two
nontrivial relations. Their validity can be verfied using the
expressions for NLO coefficients $C_q^{(1)}$ and $C_G^{(1)}$,
calculated in
\cite{Willy2,Willy3}. According to \cite{Willy1} and taking into
account slightly different normalization convention
\footnote{My $P_{ij}^{(0)}$ are by a factor of 4 smaller than
those used in \cite{Willy2,Willy3}, and $P_{ij}$ are expanded
in powers of $\alpha/2\pi$ in this paper and in powers of
$\alpha_s/4\pi$ in \cite{Willy2,Willy3,Willy1}.},
$C_{\gamma}^{(1)}$ can be obtained from $C_G^{(1)}$
in \cite{Willy3} by
\begin{itemize}
\item replacing $T_fn_f$ by unity,
\item dropping terms including $P^{(0)}_{GG}$ and $\beta_0$, which are
absent for $C_{\gamma}^{(1)}$, and
\item replacing $P^{(0)}_{qG}$ with $k_q^{(1)}$.
\end{itemize}
Substituting $C_{\gamma}^{(1)}$ obtained in this way into
(\ref{condition}) one finds that the sum on its l.h.s. indeed
vanishes. Let me emphasize that for this cancellation the
presence of $k_q^{(1)}$ in (\ref{condition}) is vital.

For the purpose of comparing LO QCD expression for $F_2^{\gamma}$,
defined in (\ref{F2LO}), with the conventional LO formula as well
as with the data, we add to it the QED contribution
(\ref{FQED}):
\begin{equation}
\frac{1}{x}\left(F_{2,\mathrm{QED}}^{\gamma}+
                 F_{2,\mathrm{LO}}^{\gamma}\right)=
q_{\mathrm{QED}}+q_{\mathrm{QCD}}+
\frac{\alpha}{2\pi}C_{\gamma}^{(0)}+
\frac{\alpha_s}{2\pi}C_q^{(1)}q_{\mathrm{QED}}
+\frac{\alpha}{2\pi}\frac{\alpha_s}{2\pi}C_{\gamma}^{(1)}.
\label{FQEDQCD}
\end{equation}
Recalling the conventional formulation of the LO QCD approximation
of $F_2^{\gamma}$
\cite{VogtinFreiburg}
\begin{equation}
\frac{1}{x}F^{\gamma}_{2,\mathrm{LO}}(x,Q^2)=q(x,Q),
\label{wrongLO}
\end{equation}
where $q(x,M)$ satisfies the evolution equation that includes
$k_q^{(0)}$ and $P_{qq}^{(0)}$ splitting functions only, we see
that it differs from (\ref{FQEDQCD})
\begin{itemize}
\item by the absence of the contributions of photonic coefficient
functions $C_{\gamma}^{(0)}$ and $C_{\gamma}^{(1)}$,
\item by the absence of the convolution of quark coefficient
function $C_q^{(1)}$ with $q_{\mathrm{QED}}$, and
\item by the fact $k_q^{(1)}$ is not
included in the evolution equation for $q(M)$.
\end{itemize}
All these differences are substantial, but as all necessary
quantities are available, there is no obstacle to performing
complete LO QCD analysis using formula (\ref{FQEDQCD}) with
$q_{\mathrm{QCD}}$ given by the formula (\ref{QLO}) of the
next subsection.

On the other hand, as neither $k_q^{(2)}$ nor $C_{\gamma}^{(2)}$
are known, a complete NLO QCD analysis of $F_2^{\gamma}$ is
currently impossible to perform.
Within the conventional approach the terms in (\ref{F2LO})
proportional to $C_{\gamma}^{(0)}$ and $C_q^{(1)}$ are part of
the NLO expression
\begin{equation}
\frac{1}{x}F_{2,\mathrm{NLO}}^{\gamma}=q
+\frac{\alpha_s}{2\pi}C_q^{(1)}q+
\frac{\alpha}{2\pi}C_{\gamma}^{(0)},
\label{wrongNLO}
\end{equation}
but the photonic coefficient function $C_{\gamma}^{(1)}$, though
known, is not used even at the NLO. As we shall see in Section 6,
its contribution is numerically comparable to that of
$C_{\gamma}^{(0)}$!

\subsection{Explicit form of the pointlike solution
${\mathbf{q_{\mathrm{NS}}(n,M,M_0)}}$}
The inhomogeneous evolution equation for nonsinglet quark distribution
function is technically particularly simple to solve in the case
$\beta_i=0,i\ge 1$. As none of the conclusions of this paper depends
in any essential way on nonzero value of $\beta_1$ and higher order
coefficients $\beta_i;i\ge 2$ can be set to zero by the choice of
an appropriate renormalization scheme, I will set $\beta_i=0;i\ge 1$
thorought the rest of this paper.
Under this simplifying assumption
$\mathrm{d}q(M)/\mathrm{d}\ln M^2=(-\beta_0/4\pi)\mathrm{d}q/
\mathrm{d}\alpha_s$
and the inhomegeneous evolution equation can be rewritten as
\begin{eqnarray}
q'(y) & \equiv &
\frac{\mathrm{d}q(y)}{\mathrm{d}y}=  -\frac{4\pi}{\beta_0}
\frac{1}{\alpha_s^2}\frac{\mathrm{d}q(M)}{\mathrm{d}\ln M^2}=\label{l1}
\nonumber \\
 & & -
\left[\frac{2\alpha k_q^{(0)}}{\beta_0}\frac{1}{y^2}+
\frac{\alpha k_q^{(1)}}
{\pi\beta_0}\frac{1}{y}+\frac{\alpha k^{(2)}}{2\pi^2\beta_0}+\cdots
\right]-
\left[\frac{2P^{(0)}_{qq}}{\beta_0}\frac{1}{y}+
\frac{P^{(1)}_{qq}}{\pi\beta_0}
+\frac{P^{(2)}_{qq}}{2\pi^2\beta_0}y+\cdots
\right]q,\label{qprime}\nonumber\\
& = & -Q(y)-P(y)q(y)\label{qdy}
\end{eqnarray}
where
\begin{eqnarray}
P(y) & = &
\frac{2P^{(0)}_{qq}}{\beta_0}\frac{1}{y}+
\frac{P^{(1)}_{qq}}{\pi\beta_0}
+\frac{P^{(2)}_{qq}}{2\pi^2\beta_0}y+\cdots
=\sum_{i=1}^{\infty}p_iy^{i-2},\label{PP}\\
Q(y) & = &
\frac{2\alpha k^{(0)}}{\beta_0}\frac{1}{y^2}+\frac{\alpha k^{(1)}}
{\pi\beta_0}\frac{1}{y}+\frac{\alpha k^{(2)}}{2\pi^2\beta_0}+\cdots
=\sum_{i=1}^{\infty}q_i y^{i-3}.
\label{QQ}
\end{eqnarray}
This standard inhomogeneous differential equation of the type
\begin{equation}
q'(y)+P(y)q(y)+Q(y)=0
\label{difequ}
\end{equation}
\FIGURE { \epsfig{file=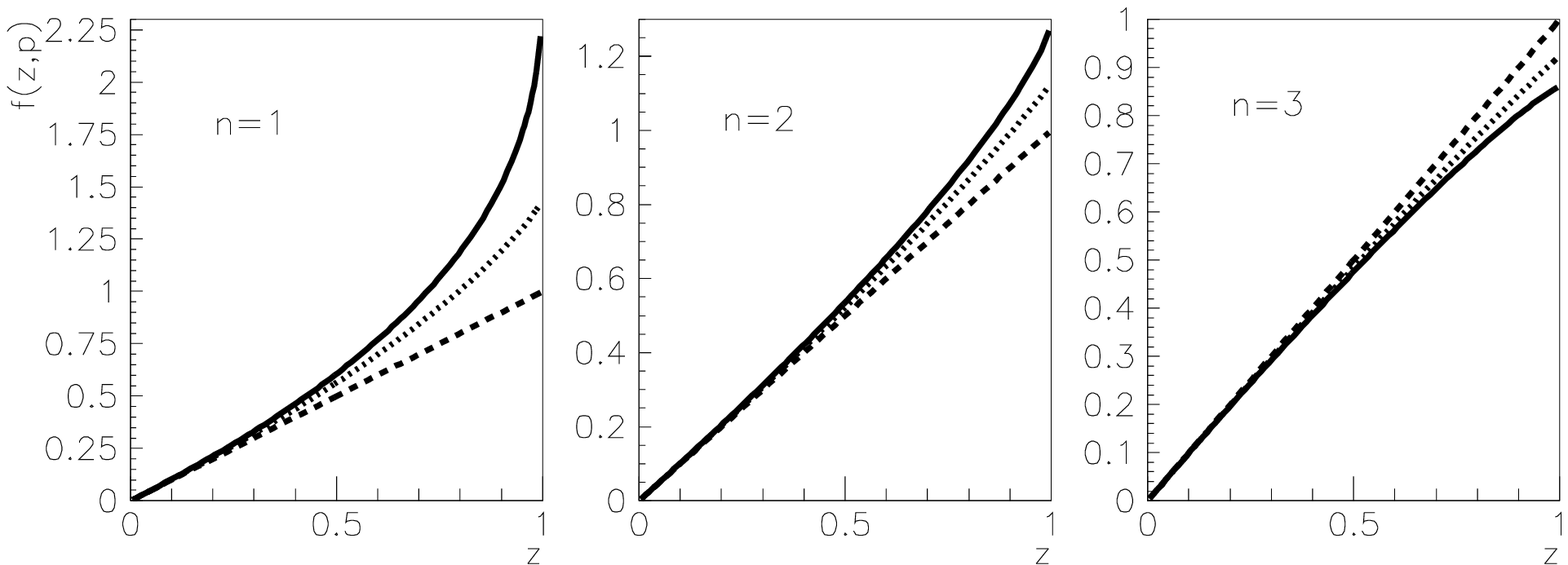,width=\textwidth}
\caption{The functions $V(z,p)$ (solid) and $W(z,p)$ (dotted),
where $p=-2P^{(0)}_{qq}(n)/\beta_0$, evaluated for $n=1,3,6$
and compared to $f(z)=z$ (dashed lines).}
\label{fplots}}
has a general solution in the form
\begin{equation}
q(y,y_0)=\mathrm{e}^{-\int_{y_0}^{y}P(u)\mathrm{d}u}
\left [q_0-\int_{y_0}^{y}Q(v)\mathrm{e}^{\int_{y_0}^{v}P(u)\mathrm{d}u}
\right],
\label{gensol}
\end{equation}
where $q_0=q(y_0)=q(M_0)$ specifies the initial condition imposed on the
solution of (\ref{gensol}). Our pointlike solutions correspond to
$q_0=0$ and we thus have
\begin{equation}
q(y,y_0)=-\mathrm{e}^{-\int_{y_0}^{y}P(u)\mathrm{d}u}
\int_{y_0}^{y}Q(v)\mathrm{e}^{\int_{y_0}^{v}P(u)\mathrm{d}u}=-
\int_{y_0}^{y}Q(v)\mathrm{e}^{\int_{y}^{v}P(u)\mathrm{d}u},
\label{pointsol}
\end{equation}
For further discussion it is useful to introduce the following functions
\begin{eqnarray}
V(z,p)& \equiv & \frac{1-(1-z)^p}{p}= \frac{(1-(y/y_0)^p)}{p},
\label{Vp}\\
W(z,p) & \equiv &\frac{-2}{p(1+p)z}\left[1-(1-z)^{1+p}-
(1+p)z\right],
\label{wdef}
\end{eqnarray}
where
\begin{equation}
z\equiv y\left(\frac{1}{y}-\frac{1}{y_0}\right)=yA=\alpha_s(M)
\frac{\beta_0}{4\pi}L_M=
\frac{\ln\frac{M^2}{M_0^2}}{\ln\frac{M^2}{\Lambda^2}}
;~~L_M\equiv \ln\frac{M^2}{M_0^2},
\label{zA}
\end{equation}
which both behave for $z\rightarrow 0$ and all $p$ like $f(z)=z$.
Their shapes are shown, for three values of $p$ corresponding
to $n=1,3,6$, in Fig. \ref{fplots}.
For fixed $M^2,M_0^2$ the region $z\rightarrow 0$ corresponds to
the limit $\alpha_s=y\rightarrow 0$. Inserting into
(\ref{pointsol}) the expansions (\ref{PP}) and (\ref{QQ}) and
retaining in both cases first three terms we get
\begin{eqnarray}
\lefteqn{q(y,y_0)=\mathrm{e}^{-\left(p_2y+\frac{1}{2}p_3y^2\right)}} & &
\label{ss1}\nonumber\\
& &
\left[~~~~q_1V(z,1-p_1)\frac{1}{y}~+~~~~~~~q_1p_2V(z,-p_1)~
+~~\frac{q_1(p_2^2+p_3)}{2}V(z,-1-p_1)y\right.
\label{ss2}\nonumber\\
& &
+\left. ~~~~~~q_2V(z,-p_1)~~~+~q_2p_2V(z,-1-p_1)y~+~
\frac{q_2(p_2^2+p_3)}{2}V(z,-2-p_1)y^2\right.
\label{ss3}\nonumber\\
& &
+\left. ~q_3V(z,-1-p_1)y~+~q_3p_2V(z,-2-p_1)y^2+
\frac{q_3(p_2^2+p_3)}{2}V(z,-3-p_1)y^3\right]
\label{qexplicit}
\end{eqnarray}
Expanding further the exponential prefactor in (\ref{qexplicit}),
adding and subtracting the purely QED contribution (\ref{qQED}) and
grouping together term that stand by same power of $y=\alpha_s(M)$
we get, keeping terms up to $y^2=\alpha_s^2$,
\begin{eqnarray}
\lefteqn{
q(y,y_0)=\frac{\alpha}{2\pi}k_q^{(0)}\ln\frac{M^2}{M_0^2}} & &
\label{QEDpart}\\ & & +\left[\frac{q_{1}}{y}V(z,1-p_1)-
\frac{\alpha}{2\pi}k^{(0)}\ln\frac{M^2}{M_0^2}+
q_2V(z,-p_1)\right]\label{LOpart}\\ & &
+\left[q_1p_2\frac{V(z,-p_1)-V(z,1-p_1)}{y}+q_3V(z,-1-p_1)-
q_2p_2V(z,-p_1) \right]y. \label{NLOpart}
\end{eqnarray}
The first line in the above expression defines the purely QED
part, $q_{\mathrm{QED}}$, the second line the LO approximation,
$q_{\mathrm{LO}}$, and the sum of (\ref{LOpart}) and (\ref{NLOpart})
$q_{\mathrm{NLO}}$. Note that the difference $(V(-p_1)-V(1-p_1))/y$
behaves at small $y$ as $\frac{1}{2}A^2y=O(y)$, i.e. in the same
way as $V(z,p)\approx z=yA=O(y)$ and therefore contributes to
$q_{\mathrm{NLO}}$.

\subsubsection{QED part}
The logarithmic rise of $F_2^{\gamma}(x,Q^2)$ with $Q^2$ results
from integration of transverse momenta (virtualities) of
quarks/antiquarks produced in the primary QED vertex
$\gamma\rightarrow q\overline{q}$ and has therefore
nothing to do with QCD. The scale $M_0$, which from the point of
view of mathematics defines the scale at which the initial
conditions on (\ref{difequ}) are imposed, has a clear physical
meaning: it separates the region of transverse
momenta (virtualities) where nonperturbative effects are
dominant \cite{FKP}. Note that although $q_{\mathrm{QED}}$ and
$C_{\gamma}^{(0)}$ separately
depend on the factorization scale $M$, their sum
$F_{2,\mathrm{QED}}^{\gamma}$ does not!
The purely QED part (\ref{FQED}) of $F_2^{\gamma}$
plays similar role as the purely
QED prefactor $3\sum e_i^2$ in (\ref{Rlarge}). We can either subtract
it from data or include it in theoretical analyses. The second option,
i.e. considering the full sum $q(x,M)$ in (\ref{qQEDQCD}) as the basic
quantity, is preferable because the QCD part
of $q(M)$ alone, i.e. $q_{\mathrm{QCD}}$, is negative
\footnote{The fact that QCD correction to $q$ is negative is nothing
extraordinary. The same holds, for instance, for QCD corrections to
Gross--Llewellyn--Smith sum rule.}.
In the following we shall analyze the
basic features of $F_{2,\mathrm{QCD}}^{\gamma}$ and $q_{\mathrm{QCD}}$
but keep in mind that for proper physical interpretation the purely
QED parts $F_{2,\mathrm{QED}}^{\gamma}$ and $q_{\mathrm{QED}}$ have to
be included as well.

\subsubsection{QCD part: leading order}
In terms of functions $V(z,p)$ and $W(z,p)$ the LO expression for
$q_{\mathrm{QCD}}$ reads
\begin{equation}
q_{\mathrm{LO}}(M^2)=
\frac{\alpha}{2\pi}\frac{k_q^{(0)}P_{qq}^{(0)}}{\beta_0}
\ln\frac{M^2}{M^2_0}
W(z,-p_{1})+\frac{\alpha}{2\pi}\frac{2k_q^{(1)}}{\beta_0}V(z,-p_{1}).
\label{QLO}
\end{equation}
The first term in (\ref{QLO}) coincides with the LO expression
(\ref{generalpointlike}) of the conventional approach from which the
purely QED part (\ref{qQED}) was subtracted. The
importance of the second term in (\ref{QLO}), which is absent in the
conventional approach, will be discussed in the next Section.
The order of (\ref{QLO}) is determined by its behaviour in the limit
$\alpha_s\rightarrow 0$. In investigating this behaviour all external
kinematical variables, as well as factorization scales must be
kept fixed and the approach $\alpha_s\rightarrow 0$ realized by
sending $\Lambda_{\mathrm{QCD}}\rightarrow 0$. In our case this means
keeping $Q^2,M^2$ and $M_0^2$ fixed and sending $z\rightarrow 0$.
In this limit both terms in (\ref{QLO})
behave in the same way and we get
\begin{equation}
q_{\mathrm{LO}}(M) \approx \left[\frac{q_1p_1A^2}{2}+q_2A\right]
\alpha_s(M)=\frac{\alpha}{2\pi}\frac{\alpha_s(M)}{2\pi}
\left[\frac{k_q^{(0)}P^{(0)}_{qq}}{2}L_M^2+
k_q^{(1)}L_M\right],
\label{F2smallyfinal}
\end{equation}
Eq. (\ref{F2smallyfinal}) is closely analogous to standard LO
expression for $r(Q)$ in (\ref{rsmall}). It is easy to check that
(\ref{F2smallyfinal}) satisfies (\ref{modified}) where only the
first term on its r.h.s. is taken into account.

The behaviour of $q_{\mathrm{LO}}(x,M)$ at large scales $M$ is
determined by (\ref{QLO}) in the limit $z\rightarrow 1$. Taking
into account that $V(1,p)=1/p$ and $W(1,p)=2/(1+p)$ we find
\begin{equation}
q_{\mathrm{LO}}(x,M)\rightarrow
\frac{\alpha}{2\pi}k_q^{(0)}\frac{2P_{qq}^{(0)}}{\beta_0}
\frac{1}{1-2P^{(0)}_{qq}/\beta_0}L_M
+\frac{\alpha}{2\pi}\frac{k_q^{(1)}}{-P^{(0)}_{qq}}.
\label{zto1}
\end{equation}
The first term in (\ref{zto1}) is dominant as
$M\rightarrow\infty$, but as we shall see in the next
Section the second one, absent in conventional LO analysis,
is actually numerically more important in most of the currently
accessible kinematical region.

\subsubsection{QCD part: next--to--leading order}
The expression for $q_{\mathrm{NLO}}(M^2)$, given by
the sum of (\ref{LOpart}) and (\ref{NLOpart}),
behaves at small $y$ as
\begin{eqnarray}
\lefteqn{
q_{\mathrm{NLO}}(M^2) \approx \left[\frac{q_{1}p_{1}A^2}{2}+
q_{2}A\right]\alpha_s(M)+
\left[\frac{q_1p_2A^2}{2}+(q_3-q_2p_2)A\right]\alpha_s(M)^2} & &
\label{L1}\\
 &= & \frac{\alpha}{2\pi}\left\{
\frac{\alpha_s(M)}{2\pi}
\left[\frac{k_q^{(0)}P^{(0)}_{qq}}{2}L_M^2+k_q^{(1)}L_M
\right]+\left(\frac{\alpha_s(M)}{2\pi}\right)^2
\left[k^{(0)}_qP^{(1)}_{qq}L_M^2+
\left(k^{(2)}_q-\frac{2k_q^{(1)}P_{qq}^{(1)}}{\beta_0}\right)L_M
\right]\right\}
\label{NLOsmally} \nonumber
\end{eqnarray}
and involves in addition
to quantities entering the LO formula (\ref{QLO}), i.e.
$k^{(0)}_q, k_q^{(1)}$ and $P_{qq}^{(0)}$ also the $O(\alpha_s^2)$
inhomogeneous splitting function $k_1^{(2)}$ and $O(\alpha_s^2)$
homogeneous splitting function $P^{(1)}_{qq}$. Whereas the latter is
known, the former is not, thereby preventing the evaluation of
$q_{\mathrm{NLO}}$. In addition to this obstacle, a  complete NLO
analysis of $F_2^{\gamma}$ requires also the knowledge of
$O(\alpha\alpha_s^2)$ photonic coefficient function $C_{\gamma}^{(2)}$,
which is also so far unknown. Consequently, a complete NLO QCD
analysis of $F_2^{\gamma}$ is currently impossible to perform.

\section{Phenomenological implications}
In this Section I will demonstrate numerical importance of the
contributions to $F_{2,\mathrm{LO}}^{\gamma}$ resulting from the
inclusion of the quantities that are omitted in standard LO
analysis: the inhomogeneous splitting function $k_q^{(1)}$,
photonic coefficient functions $C_{\gamma}^{(0)}$ and
$C_{\gamma}^{(1)}$ and quark
coefficient function $C_q^{(1)}$. All quantitative considerations
concern the LO QCD analysis of ${\mathrm {F_2^{\gamma}}}$ in the
nonsinglet channel only, assuming, for the sake of technical
simplicity, $\beta_1=0$. Any realistic analysis of experimental
data will require inclusion of effects of nonzero value of
$\beta_1$, as well as extension of the formalism to the singlet
channel and incorporation of contributions of hadronic parts of
PDF. The work on these problems is in progress. Nevertheless, the
quantitative impact of some of the effects discussed in this
Section on the LO QCD analysis of $F_2^{\gamma}$ is so large that
we may expect them to affect significantly the complete LO
analysis of $F_2^{\gamma}$ as well.
\FIGURE{ \epsfig{file=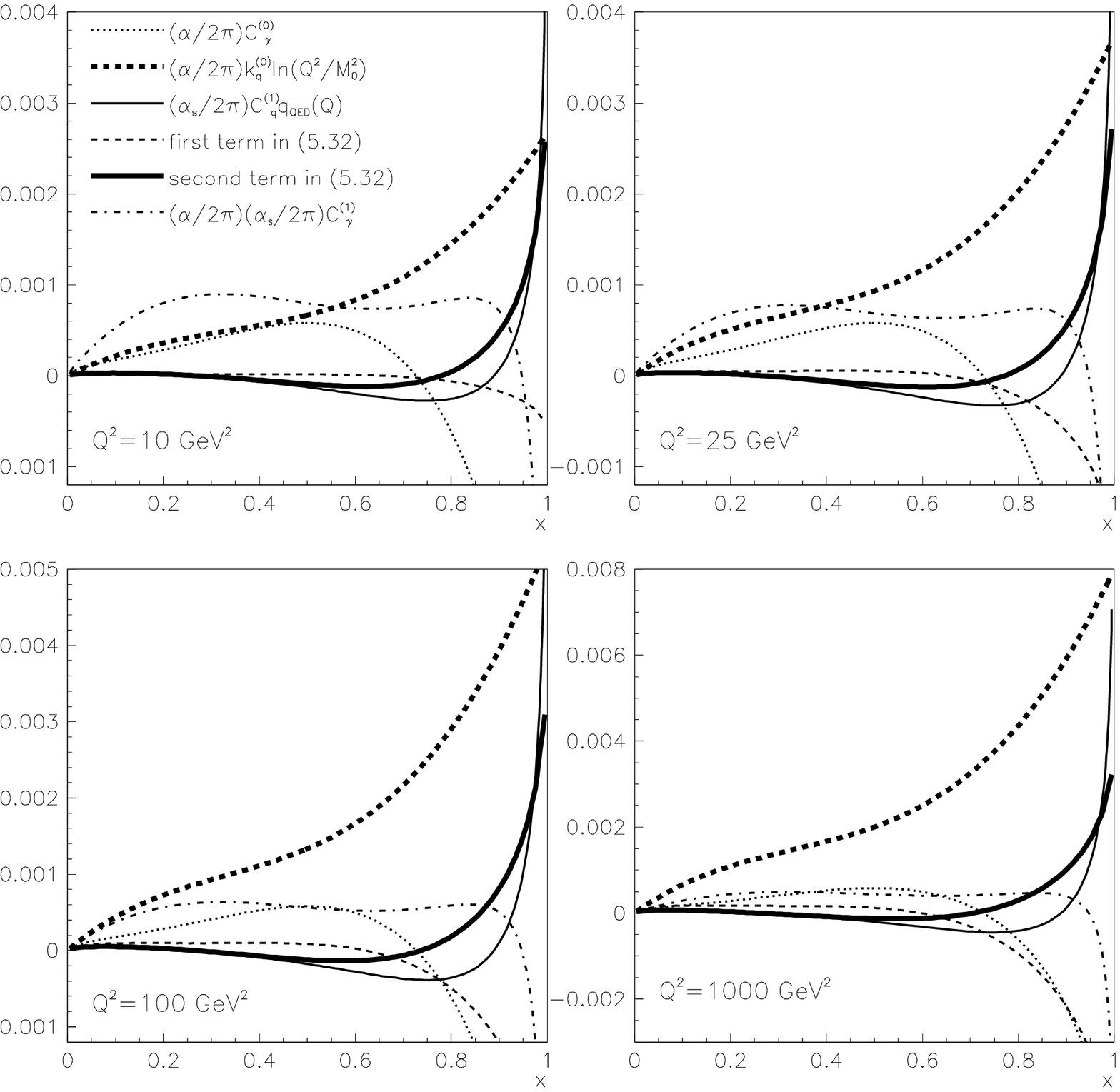,width=\textwidth}
\caption{Comparison of individual contributions
to $F_2^{\gamma}$ as described in the text.}
\label{c2fig} }

\subsection{The effects of ${\mathbf{C_{\gamma}^{(0)}}}$}
In all consideration of this Section I will take for
$C_{\gamma}^{(0)}$ the standard expression (\ref{C0}), evaluated
for $M^2=Q^2$.
The presence of the lower bound $M_0^2$ on quark virtuality
complicates, however, the situation and I will therefore
discuss this point in more detail in subsection 6.6.

\subsection{The effects of ${\mathbf{C_{\gamma}^{(1)}}}$}
For illustration of the numerical importance of the $O(\alpha_s)$
photonic coefficient function $C_{\gamma}^{(1)}$, its contribution
to LO analysis of $F_2^{\gamma}$ in $\overline{\mathrm
{MS}}+$MS scheme is compared in Fig. \ref{c2fig} to those of
$C_{\gamma}^{(0)}(x)$, the QED formula (\ref{qQED}), as well as
other QCD contributions discussed in this Section.
The comparison is performed for $M^2=10,25,100$ and $1000$
GeV$^2$. Contrary to the contribution of $C_{\gamma}^{(0)}$,
$C_{\gamma}^{(1)}$ enters the expression for $F_2^{\gamma}$ multiplied
by $\alpha_s(Q)$, and therefore decreases with increasing
$Q^2$ until it becomes negligible with respect to the former
as $Q\rightarrow \infty$. However, one would have to go to extremely
large $Q^2$, inaccessible in current experiments, for this
dominance of $C_{\gamma}^{(0)}$ to be a good approximation. Fig.
 \ref{c2fig} shows that in the whole
currently accessible kinematical region of $Q^2$ and for $x$ up to
about $0.85$, the term proportional to $C_{\gamma}^{(1)}$ provides
numerically the most important $O(\alpha_s)$ correction to QED
formula (\ref{FQED}).

\subsection{The effects of ${\mathbf{C_{q}^{(1)}}}$}
Performing the convolution $C_q^{(1)}\otimes q_{\mathrm{QED}}$, implied by
(\ref{FQEDQCD}), where
\begin{equation}
C_q^{(1)}=C_F\left(\frac{9+5x}{4}-\frac{1+x^2}{1-x}\ln x-\frac{3}{4}
\frac{1+x^2}{[1-x]_+}+\left(1+x^2\right)\left[\frac{\ln(1-x)}{1-x}
\right]_+-\left(\frac{9}{2}-\frac{\pi^2}{3}\right)\right)
\label{Cqq}
\end{equation}
is straightforward and leads to the thin solid curves in Fig.
\ref{c2fig}. This contribution has qualitatively the same shape as
that of $k_q^{(1)}$, discussed below, and is vital
in the region close to $x=1$.

\subsection{The effects of ${\mathbf{k_q^{(1)}}}$}
\FIGURE { \epsfig{file=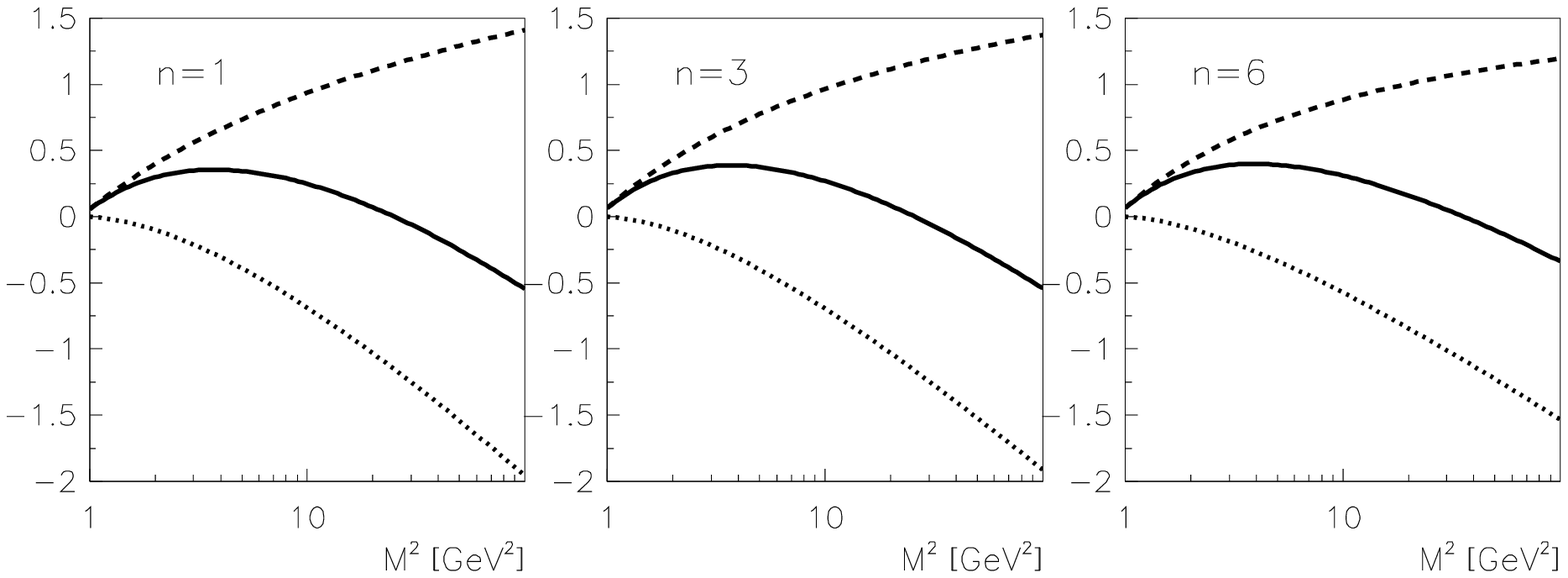,width=12cm} \caption{The LO
expression (\ref{QLO}), multiplied by $2\beta_0\pi/\alpha$, as a
function of $M^2$ for three moments $n=1,3,6$, $M_0=1$ GeV and
$\Lambda=0.2$ GeV. The solid line shows the full result, the
dotted and dashed ones correspond to the first and second term in
(\ref{QLO}) plotted separately.}
\label{qloplots}}
\FIGURE{ \epsfig{file=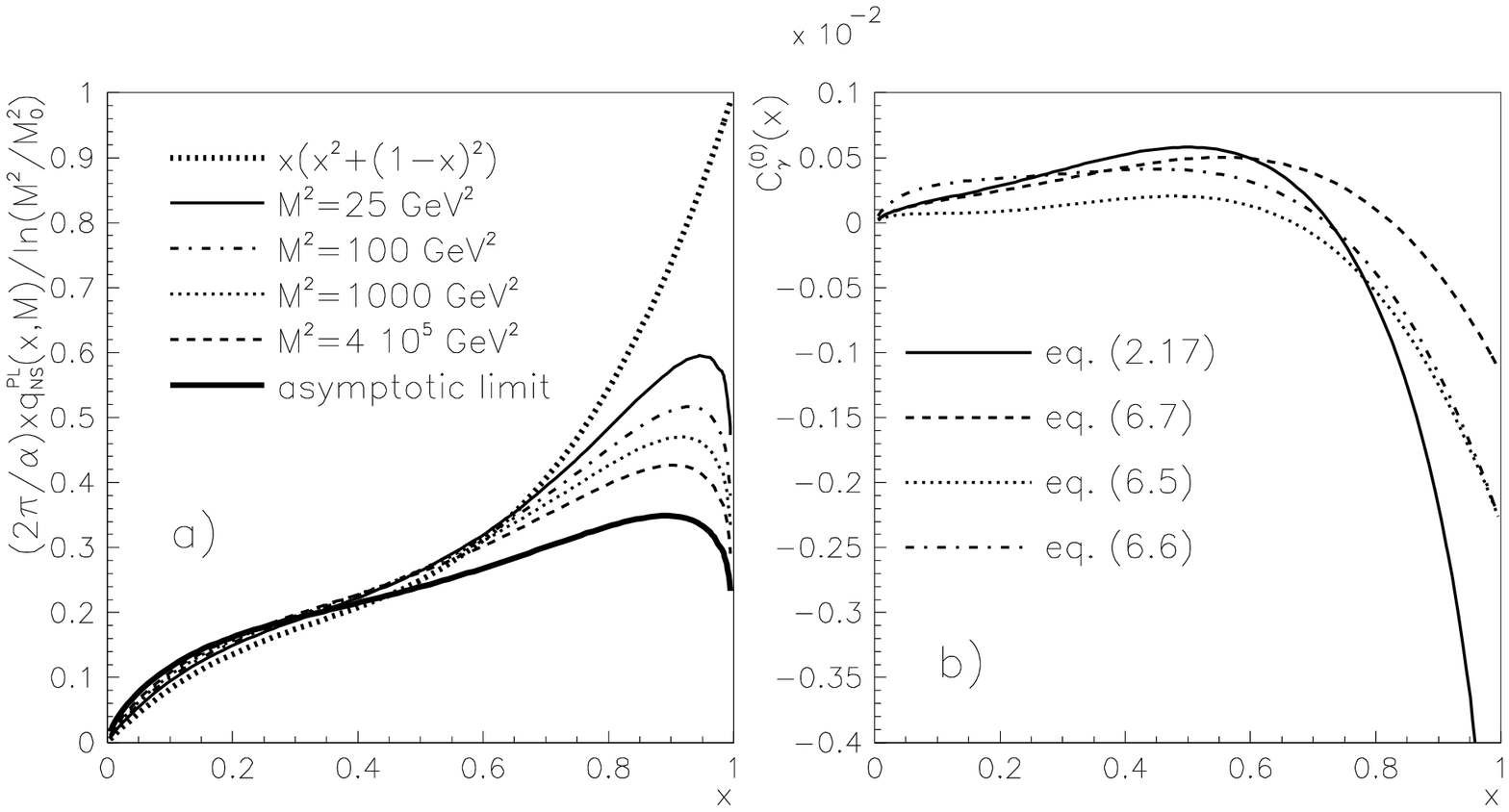,width=\textwidth}
\caption{Illustration of the approach to the asymptotic pointlike
expression (\ref{asymptotic}) (a), and the comparison of three
expressions for $C^{(0)}_{\gamma}(x)$ discussed in the text (b).}
\label{C0fig} }
\noindent
At asymptotically large values of $M^2$
the first term on the r.h.s. of (\ref{QLO}), which is the only one
included in standard LO analysis of $F_2^{\gamma}$, dominates due
to the
presence of ``large log'' $\ln(M^2/M_0^2)$
but again one would have to go to very large $M^2$, inaccessible in
current  experiments
\footnote{For $M_0\doteq 1$ GeV and
currently accessible  $M^2<200$ GeV$^2$ the ``large log''
$\ln (M^2/M_0^2)\lesssim 5$, hardly a large number.},
to see this dominance numerically. This is
illustrated in Fig. \ref{qloplots}, where $q_{\mathrm{LO}}(M)$,
as defined in (\ref{QLO}), is
plotted as a function of $M^2$ for three moments
$n=1,3,6$ and together with separate contributions of the
first and second terms in (\ref{QLO}). We see that for $M^2\le
100$ GeV$^2$ the effect of including the latter almost cancels the
negative contribution of the term appearing in the conventional
analysis. Moreover, converting (\ref{QLO}) into the $x$-space by
means of inverse Mellin transformation we find (see Fig. \ref{c2fig})
that the negative
contribution of the first term in (\ref{QLO})
to moments of $q_{\mathrm{LO}}(n,M)$ comes mostly from the region
close to
$x=1$. Below $x\doteq 0.75$ the genuine QCD effects described by this
part of (\ref{QLO}) are tiny and smaller than those
described by the second term, proportional to the inhomogeneous
splitting
function $k_q^{(1)}$! The latter gives small negative contribution
to $q_{\mathrm{LO}}(x,M)$ up to $x\doteq 0.7$ and large positive one
above that
value. Figs. \ref{c2fig} and \ref{qloplots} show convincingly that in
the whole region of $Q^2$ and in most of the range of $x$ accessible
experimentally, the second
term in (\ref{QLO}) is numerically
more important than the one included in standard LO analyses.
Only for $x$ close to $x=1$ are both terms in (\ref{QLO}) comparable
and more or less cancelling each other.
The tiny effect of QCD corrections described by the first part of
(\ref{QLO}) reflects the fact that in the region
$x\lesssim 0.65 $ its contribution to scaling
violations of the sum $q_{\mathrm{QED}}+q_{\mathrm{QCD}}$,
plotted in Fig. \ref{C0fig}a, are very small.
\FIGURE { \epsfig{file=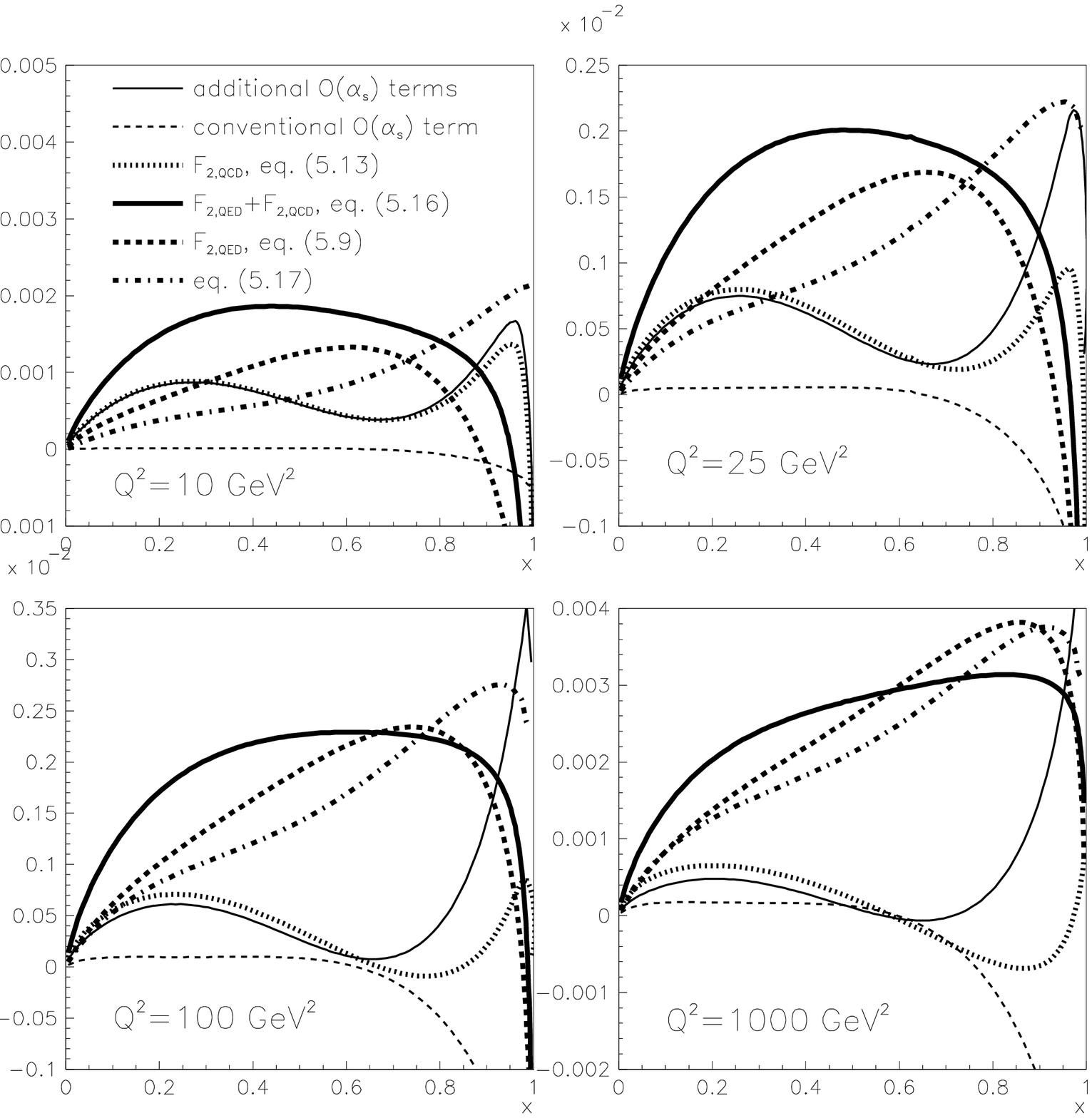,width=\textwidth}
\caption{The sum of contributions to LO QCD expression for
$F_2^{\gamma}$ included in this analysis, but omitted in the
conventional ones, compared to the the effects of $P^{(0)}_{qq}$
as well as to the QED formula (\ref{FQED}), complete LO expression
(\ref{FQEDQCD}), and standard LO formula (\ref{wrongLO}).}
\label{sums}}

\subsection{Summary of QCD contributions at the LO}
In Fig. \ref{sums} the sum of contributions to LO analysis of
$F_2^{\gamma}$ coming from inclusion of $C_{\gamma}^{(1)},
C_q^{(1)}$ and $k_q^{(1)}$ is compared to that given by the first
term in (\ref{QLO}), the purely QED contribution (\ref{FQED}), the
complete LO QCD contribution (\ref{F2LO}), as well as full
QED$+$QCD expressions in my (\ref{FQEDQCD}) and standard
(\ref{wrongLO}) approaches, in the latter including the
contribution of $C_{\gamma}^{(0)}$ as well. In all calculations
$M^2=Q^2$. The difference between the solid and dash--dotted
curves, which represent my and standard expressions for
$F_{2,\mathrm{LO}}^{\gamma}$, is large and
persists for all experimentally accessible values of $Q^2$. This
is caused by the fact that in the whole kinematical region of
$Q^2$ and for most of the region of $x$ accessible in current
experiments, the sum of the three additional contributions
included in the LO analysis proposed in this paper, is much larger
than the contribution of the term generated by $P^{(0)}_{qq}$. But
even for $x$ close to $x=1$ these additional terms are important
as they help cancel in part the negative contribution of
$C_{\gamma}^{(0)}$.

\subsection{Dispensing with the
DIS$_{\gamma}$ factorization scheme}
The quantitative comparison in Fig. \ref{sums}
reveals another interesting
difference: the term $\ln(1-x)$ in (\ref{C0}) is much less troubling
within my formulation than within the standard one. The reason
is simple: the contributions of additional terms compensate large
part of the negative contribution of $C_{\gamma}^{(0)}$ at large
$x$ and shifts the region where the complete LO expression turns
negative much closer to $x=1$.

A simple explanation of the origins of this term is provided within
the parton model derivation \cite{factor} of the expression
(\ref{C0}) in which the parallel singularity associated with the
splitting
$\gamma\rightarrow q\overline{q}$ is regularized by means of quark
masses and the troubling term $\ln(1-x)$
comes from the lower bound on the quark virtuality $\tau$
(see Fig. \ref{figpl}) in collinear kinematics
\begin{equation}
\tau_{\mathrm {min}}^{\mathrm {coll}}\equiv \frac{m_q^2}{1-x}\le
\tau\le
\tau_{\mathrm {max}}^{\mathrm {coll}}\equiv \frac{Q^2}{x}.
\label{bounds}
\end{equation}
Performing the integration over $\tau$ in the limits
$\tau_{\mathrm {min}}^{\mathrm {coll}}\le \tau\le M^2$,
as prescribed by the definition (\ref{dressed}) of PDF of the
real photon, the primary QED splitting
$\gamma\rightarrow q\overline{q}$ leads to \cite{factor}
\begin{equation}
f(x)\ln\left(\frac{M^2}
{\tau_{\mathrm{min}}^{\mathrm{coll}}}\right)
=f(x)\ln\left(\frac{M^2(1-x)}{m_q^2}\right)
=f(x)\ln\left(\frac{M^2}{m_q^2}\right)+f(x)\ln(1-x),
\label{logterm}
\end{equation}
where $f(x)=x^2+(1-x)^2$.
Normally, the term proportional to $\ln(M^2/m_q^2)$ is included
in the quark distribution function $q(x,M)$, whereas the ``constant''
part of (\ref{logterm}), i.e. $f(x)\ln(1-x)$ goes into the
coefficient function $C_{\gamma}^{(0)}$.
Note that we cannot set $m_q=0$ in the logarithmic term.
The complete expression for nonlogarithmic term of (\ref{C0})
includes, beside the term $f(x)\ln(1-x)$,
also terms coming from the upper bound in (\ref{bounds}) as well
as from lower bound in integrals over the terms $m_q^2/\tau^2$,
or $1/s$. In these latter cases the dependence on $m_q$ enters
$F_2^{\gamma}(x,Q^2)$ through the multiplicative factor
\begin{equation}
c\equiv 1-\frac{\tau_{\mathrm{min}}^{\mathrm{coll}}}{M^2}=
1-\frac{m_q^2}{M^2(1-x)},
\label{cfactor}
\end{equation}
where we can set $m_q=0$ in (\ref{cfactor}). Moreover, as
the ratio $\tau_{\mathrm{min}}^{\mathrm{coll}}/M^2$ is formally
of power correction type, we can neglect it even for $m_q>0$.
The term $\ln(1-x)$ in $C^{(0)}_{\gamma}$ causes
problems in phenomenological parameterizations primarily because
it appears there decoupled from the value of the quark mass $m_q$
with which it originally entered the
expression (\ref{bounds}) for $\tau_{\mathrm{min}}$ and thus
persists there even in the limit $m_q\rightarrow 0$, or when
the quark mass is replaced by the initial scale $M_0$.

In the presence of the lower cutt--off $M_0^2$ on the integration
over the quark virtualities
$\tau_{\mathrm{min}}^{\mathrm{coll}}$ in (\ref{bounds})
should perhaps be replaced with
$\tau_0\equiv\mathrm{max}(m_q^2/(1-x),M_0^2)$. Sending
$m_q\rightarrow 0$, but even for all realistic values of light quark
masses and accessible $x$, we get $\tau_0=M_0^2$. This implies
the replacement $\ln((1-x)/x)\rightarrow \ln(1/x)$
in $C_{\gamma}^{(0)}$, thereby
removing most of the practical problems with the term $\ln(1-x)$
in (\ref{C0}).

Beside $\ln(1-x)$ there is another term in $C_{\gamma}^{(0)}$ that
comes from nonzero quark mass, namely the last ``$+1$'' in the
nonlogarithmic part $8x(1-x)-1=8x(1-x)-2+1$. Discarding also this
remaining trace of $m_q$ we get, setting $M^2=Q^2$,
\begin{equation}
C_{\gamma}^{(0)}(x,1)=\left[x^2+(1-x)^2\right]
\ln\frac{1}{x}+8x(1-x)-2.
\label{C0my}
\end{equation}
This expression is close to, but not identical QED formula
for the constant part of $F_2^{\gamma}(x,Q^2)$ obtained for massless
quarks coupled photon with nonzero virtuality $P^2>0$ \cite{russians}
\begin{equation}
C_{\gamma}^{(0)}(x,1)=\left[x^2+(1-x)^2\right]
\ln\frac{1}{x^2}+8x(1-x)-2.
\label{C0P2}
\end{equation}
The only difference between (\ref{C0my}) and (\ref{C0P2}), i.e. the
additional $1/x$ in the logarithm $\ln(1/x^2)$ with respect to
(\ref{C0my}) reflects the fact that for $m_q=0$ and $P^2>0$ the lower
bound $\tau_{\mathrm{min}}^{\mathrm{coll}}=xP^2$.
The expression (\ref{C0my}) for $C_{\gamma}^{(0)}$  is also close to
that used for different reasons by Schuler and Sj\"{o}strand in their
SaS1M and SaS2M parameterizations
\begin{equation}
C_{\gamma}^{(0)}(x,1)=\left[x^2+(1-x)^2\right]
\ln\frac{1}{x}+6x(1-x)-1.
\label{C0SaS}
\end{equation}
In Fig. \ref{C0fig} all three expressions (\ref{C0my}-\ref{C0SaS})
are compared to the standard one of eq. (\ref{C0}).

\section{Conclusions}
I have presented a reformulation of LO and NLO QCD analysis of
$F_2^{\gamma}$, which separates genuine QCD effects
from those due to pure QED and satisfies the basic requirement
of factorization scale invariance. This reformulation differs
substantially from the conventional LO and NLO analysis of
$F_2^{\gamma}$.

Compared to the standard formulation at the LO it requires
the inclusion of four
additional terms, proportional to $C_q^{(1)},k_q^{(1)},
C_{\gamma}^{(0)}$ and $C_{\gamma}^{(1)}$,
all of which are known. Whereas in the
standard approach the first three of them are part of the NLO
approximation, the last one, i.e. the one proportional to
$C_{\gamma}^{(1)}$, enters the standard formulation
first at the NNLO. Detailed numerical analysis of the
contributions of all these terms shows that in most of the
kinematical region accessible experimentally their sum is much
more important than the contribution of the term included in
the standard LO expression for $F_2^{\gamma}$ and induced by
$P_{qq}^{(0)}$. It is shown that in the reformulated LO QCD
analysis of $F_2^{\gamma}$ the part of $C_{\gamma}^{(0)}$
proportional to $\ln(1-x)$, which in the standard formulation
causes problems in the region $x\rightarrow 1$, is much less
troubling. All the quantitative considerations were carried
out for the pointlike part of $F_2^{\gamma}$ in the nonsinglet
channel and under the simplifying assumption $\beta_1=0$.
A realistic analysis of experimental data will necessitate the
extension of the present formulation to the pointlike part of
$F_2^{\gamma}$ in the singlet channel as well as the addition of
the hadronic components in both channels.
At the NLO a complete analysis requires the knowledge of two
so far uncalculated quantities and is therefore currently
impossible to perform.

\acknowledgments
I thank W. van Neerven, E. Laenen and S. Larin for
correspondence concerning higher order QCD calculations of photonic
coefficient and splitting functions. I have enjoyed numerous
discussion and correspondence on the subject of photon structure
with J. Field and F. Kapusta who have advocated part of what I claim
here for more than a decade.


\begin{thebibliography}{99}
\bibitem{Vogt} A. Vogt, in {\em Procedings PHOTON '97},
Egmond aan Zee, May 1997, edt. F. Ern\'{e}, World Scientific 1998
\bibitem{Stefan} S. S\"oldner--Rembold, in {\em Procedings
 of XVIII International
 Symposium on Lepton-Photon
 Physics}, Hamburg, 1997, hep-ex/9711005.
\bibitem{Richard} R. Nisius, in {\em Proceedings PHOTON '99},
Freiburg in Breisgau,
May 1999, ed. S. Soeldner--Rembold, Nucl. Phys. Proc. Supp., in press
\bibitem{factor} J. Ch\'{y}la, hep-ph/9811455
\bibitem{FKP} J.H. Field, F. Kapusta, L. Poggioli, Phys. Lett. B
{\bf 181}, 362 (1986)\\
J.H. Field, F. Kapusta, L. Poggioli, Z. Phys. C {\bf 36}, 121 (1987)\\
F. Kapusta, Z. Phys. C {\bf 42}, 225 (1989)
\bibitem{VogtinFreiburg} A. Vogt, in {\em Proceedings PHOTON '99},
Freiburg in Breisgau,
May 1999, ed. S. Soeldner--Rembold, Nucl. Phys. Proc. Supp., in press
\bibitem{smarkem1} J. Ch\'{y}la, M. Ta\v{s}evsk\'{y},
in {\em Proceedings PHOTON '99},
Freiburg in Breisgau,
May 1999, ed. S. Soeldner--Rembold, Nucl. Phys. Proc. Supp., in press,
hep-ph/9906552
\bibitem{smarkem2}
J. Ch\'{y}la, M. Ta\v{s}evsk\'{y}, in {\em Proceedings of Workshop MC
generators for HERA Physics}, Hamburg 1999, p. 239, hep-ph/9905444
\bibitem{Aurenche} P. Aurenche, J.-P. Guillet, M. Fontannaz, Z. Phys. C
{\bf 64}, 621 (1994)
\bibitem{GRVNS} M. Gl\"{u}ck, E. Reya, A. Vogt, Phys. Rev. {\bf
D45}, (1991), 3986
\bibitem{politzer} H. D. Politzer, Nucl. Phys. B {\bf 194}, 493 (1982)
\bibitem{bardeen} W. Bardeen, A. Buras, D. Duke and T. Muta, Phys. Rev D
{\bf 18}, 3998 (1978)
\bibitem{jch2} J. Ch\'{y}la, Z. Phys. C {\bf 43}, 431 (1989)
\bibitem{FP} G. Curci, W. Furmanski, R. Petronzio, Nucl. Phys. B {\bf 175},
27 (1980)
\bibitem{fontannaz} P. Aurenche, R. Baier, A. Douiri, M. Fontannaz,
D. Schiff, Nucl. Phys. B {\bf 286}, 553 (1987)\\ P. Aurenche, R.
Baier, M. Fontannaz, D. Schiff, Nucl. Phys. B {\bf 296}, 661 (1987)
\bibitem{sas1} G. Schuler, T. Sj\"{o}strand, Z. Phys. C {\bf 68}, 607
(1995)
\bibitem{witten} E. Witten, Nucl. Phys. {\bf 120}, 189 (1977)
\bibitem{BB} W. Bardeen, A. Buras, Phys. Rev. D {\bf 20}, 166 (1979)
\bibitem{jch1} J. Ch\'{y}la, Phys. Lett. B {\bf 320}, 186 (1994)
\bibitem{marco} M. Stratmann, Talk presented at the {\em Workshop on
Photon Interactions and the Photon Structure, Lund, 1998},
hep--ph/9811260
\bibitem{Willy2} E.B. Zijlstra and W.L. van Neerven,
Phys. Lett. B {\bf 273}, 476 (1991)
\bibitem{Willy3} E.B. Ziljstra and W.L. van Neerven, Nucl. Phys. B
{\bf 383}, 525 (1992)
\bibitem{Willy1} E. Laenen, S. Riemersma, J. Smith and
W.L. van Neerven, Phys. Rev. D {\bf 49}, 5753 (1994)
\bibitem{russians} A. Gorski, B.L. Ioffe, A. Yu. Khodjamirian, A. Oganesian,
Z. Phys. C {\bf 44}, 523 (1989)
\end{thebibliography}
\end{document}